\documentclass[a4paper,fleqn]{cas-dc}

\usepackage[numbers]{natbib}

\def\tsc#1{\csdef{#1}{\textsc{\lowercase{#1}}\xspace}}
\tsc{WGM}
\tsc{QE}

\usepackage{graphicx}%
\usepackage{multirow}%
\usepackage{amsmath,amssymb,amsfonts}%
\usepackage{amsthm}%
\usepackage{mathrsfs}%
\usepackage[title]{appendix}%
\usepackage{xcolor}%
\usepackage{textcomp}%
\usepackage{manyfoot}%
\usepackage{booktabs}%
\usepackage{algorithm}%
\usepackage{algorithmicx}%
\usepackage{algpseudocode}%
\usepackage{listings}%
\usepackage{verbatim}%
\usepackage{svg}%
\usepackage{subcaption}
\usepackage[normalem]{ulem}
\usepackage{makecell}
\usepackage{placeins}
\svgsetup{inkscapelatex=false}
\usepackage{multicol}
\usepackage{bm}

\begin{document}
\let\WriteBookmarks\relax
\def\floatpagepagefraction{1}
\def\textpagefraction{.001}

\shorttitle{Morphological Features of Pulmonary Arterial Networks}    

\shortauthors{Maceda, Miller, Reyes-Roza et al.}  

\title [mode = title]{Morphological Features of Murine Pulmonary Arterial Networks for Diagnosing Vascular Disease}  

\author[1]{Elliot Maceda}
\fnmark[1]
\credit{Methodology, Software, Formal Analysis, Visualization, Writing - Original Draft}

\affiliation[1]{organization={Department of Statistics, North Carolina State University},
            addressline={2311 Stinson Drive}, 
            city={Raleigh},
            postcode={27607}, 
            state={North Carolina},
            country={USA}}

\author[2]{John M Miller}[orcid=0009-0007-2429-6634]
\fnmark[1]
\credit{Methodology, Software, Formal Analysis, Visualization, Writing - Original Draft}
\affiliation[2]{organization={Department of Mathematics, North Carolina State University},
            addressline={2311 Stinson Drive}, 
            city={Raleigh},
            postcode={27607}, 
            state={North Carolina},
            country={USA}}

\author[2]{Sebastian Reyes-Roza}
\credit{Software, Formal Analysis, Writing - Original Draft}

\author[3]{Megan Miller}
\credit{Methodology, Software, Formal Analysis, Visualization, Writing - Original Draft, Writing - Review \& Editing}
\affiliation[3]{organization={Department of Applied Mathematics, Virginia Military Institute},
            addressline={319 Letcher Avenue}, 
            city={Lexington},
            postcode={24450}, 
            state={Virginia},
            country={USA}}

\author[4]{Nicholas A Hill}
\credit{Conceptualization, Methodology, Writing - Review \& Editing}
\affiliation[4]{organization={School of Mathematics and Statistics, The University of Glasgow},
            addressline={University Avenue}, 
            city={Glasgow},
            postcode={G12 8QQ}, 
            state={Scotland},
            country={UK}}

\author[2]{Radmila Sazdanovi\'c}%
\credit{Methodology, Writing - Review \& Editing}

\author[1]{Nathaniel Josephs}
\credit{Conceptualization, Methodology, Writing - Review \& Editing}

\author[2]{Mette S Olufsen}
\cormark[1]
\ead{msolufse@ncsu.edu}
\credit{Conceptualization, Methodology, Writing - Review \& Editing}
\cortext[1]{Corresponding author}

\fntext[1]{These authors contributed equally to this work.}

\begin{abstract}
Hypoxia-induced pulmonary hypertension results from prolonged time spent in a low-oxygen environment. The induced pulmonary hypertension leads to vascular remodeling, including vessel dilation and decreased compliance. This study aims to identify changes in vascular morphology in mice with pulmonary hypertension induced by hypoxia. To do so, we examine the pulmonary arterial network morphology using methods from geometry, topological data analysis, and statistical shape analysis. Tools from these fields are applied to arterial networks excised from micro-CT images of thirteen mice at four pressures. Networks from the hypertensive mice were normalized using a radius-based pruning algorithm, and analysis was conducted on the entire arterial network, as well as on a per-lobe basis. The main objective of this study was to present an adaptable pipeline for extracting features from micro-CT images to study how vascular disease affects arterial network morphology. Using this pipeline, we showed that the hypertensive mice have increased radius ratios and decreased length-to-radius ratios. In the processed networks, HPH mice had lower mean total finite degree-one persistence of the arterial point cloud and lower mean directional complexity in both the posterior and anterior directions. Persistence curves, particularly Betti curves, revealed lobe-specific differences in the left and inferior lobes. The first two shape principal component scores were also able to discriminate between control and pulmonary hypertension at lower pressures. Importance metrics for each feature are computed using a boosted decision tree, and then an SVM classification model is fit with the most informative predictors for hypertensive classification.
\end{abstract}



\maketitle

\section{Introduction}\label{sec:intro}

\noindent Vascular disease is among the leading causes of death worldwide \cite{WHO}. It can be characterized from two perspectives: (a) changes in hemodynamics (blood pressure and flow) and (b) changes in vascular morphometry. Vascular disease causes arterial and venous remodeling \cite{Li04, Renna13}, but associated morphological changes at the system or organ level are rarely used for diagnosis or treatment. Instead, most studies that aim to quantify vascular disease, such as pulmonary hypertension \cite{Johnson23}, focus on observed changes in hemodynamics. Blood flow can be measured with ultrasound or 4D Magnetic Resonance Imaging (4D-MRI), but ultrasound measures velocity rather than blood flow \cite{Blanco15}, and 4D-MRI is expensive and has low temporal resolution \cite{Stankovic14}. Although systemic systolic and diastolic blood pressure can be measured with a blood pressure cuff \cite{Saugel20}, in the pulmonary circulation, blood pressure can only be measured invasively via right-heart catheterization \cite{Johnson23}.  

This study is motivated by the increasing availability of high-resolution images, often acquired during diagnostic protocols, and by Rahaghi et al. \cite{Rahaghi16}, who examined CT images to reconstruct three-dimensional pulmonary vascular networks in patients with and without chronic thromboembolic pulmonary hypertension. Their analysis identified key biomarkers, including pruned distal vasculature and increased arterial vessel radii, that distinguished between groups. Inspired by their work, we propose extracting a set of biomarkers from arterial networks derived from murine micro-CT images obtained {\em in vitro}. We develop methodologies to extract morphometric markers using geometric, topological data analysis (TDA), and statistical shape analysis (SSA) techniques. We derive features from arterial networks extracted from control and pulmonary hypertensive mice and use statistical classification to compare features between these groups. 

While our methods can be applied to various diseases, we focus on morphological changes associated with pulmonary hypertension (PH), defined by a mean pulmonary arterial pressure (mPAP) exceeding 20 mmHg \cite{Simonneau19, Johnson23}. It is an incurable and life-threatening disease affecting around 1\% of the global population \cite{Mocumbi24}. Treatments focus on symptom management and improving quality of life \citep{PHTreatment2022}, which require early diagnosis \citep{Bazan2012}. However, since early symptoms, including shortness of breath, fatigue, chest pain, and dizziness, mimic common conditions such as asthma, chronic obstructive pulmonary disease, and anxiety, the disease is often diagnosed too late, making treatments ineffective \cite{Vachiery12, Bazan2012}.
Currently, PH must be diagnosed by mPAP measurements from right heart catheterization \cite{Wijeratne18}. However, advanced medical imaging technology has proven to be a useful and minimally invasive diagnostic tool \cite{Rahaghi16, Rengier19}. This is especially promising for diagnosing PH early since the disease is known to cause quantifiable structural changes in the pulmonary arteries \cite{Renna13, Cajigas16}. 

The idea of examining network morphology dates back more than 100 years to studies by Murray \cite{Murray26, Murray26a}, who hypothesized that, to reach all regions of the lung, the vascular network follows an optimal principle that minimizes the energy required to transport blood through the network. This principle was encoded in a radius exponent $\xi$ that links the radius of a parent vessel to the radii of two daughter vessels. Many studies have further explored this idea by optimizing the branching ratio. A recent analysis of branching ratios by Altieri Correa et al. \cite{Correa25} in multiple species found that in the pulmonary vasculature, the Murray exponent ranged between 2.01 and 3.17. Talou et al. \cite{Talou21} developed an adaptive constrained constructive optimization algorithm to construct space-filling vascular networks. However, these studies focused on healthy networks, and the relationship likely does not hold in diseased networks. To our knowledge, only a few studies have examined how the disease affects this ratio. One study by Schoenenberger et al. \cite{Schoenenberger14} showed that near the coronary bifurcations, deviations from Murray’s law are associated with plaque composition. In our preliminary studies \cite{Chambers20, Miller23}, we noted that branching ratios $\alpha$ and $\beta$ were not impacted by PH, but these studies did not investigate the effect on $\xi$. 

Another metric associated with morphological changes is the BV5/TBV ratio, the total volume of small vessels with a cross-sectional area $< 5\text{ mm}^2$ (BV5) divided by the total blood volume (TBV). In Synn et al. \cite{Synn19}, CT images of the pulmonary vasculature showed that cigarette exposure was associated with higher pulmonary blood vessel volumes, especially in the smaller peripheral vessels. This metric is only one of many feature-based quantities that can be extracted from images. Beyond the ``simple" features discussed above, many other techniques, including TDA and SSA, can help quantify morphometric differences. 

TDA is a framework for extracting shape-based features from data using standard tools from algebraic topology \cite{Edelsbrunner02}. A commonly used TDA method is persistent homology, which tracks the appearance and disappearance of topological features such as connected components, loops, and voids as a function of a varying filtration parameter. TDA is particularly effective at identifying structure in settings with noisy data, such as medical imaging \cite{Singh23}. Miller et al. \cite{Miller23} used height filtration, a tool from persistent homology, to identify differences between control and hypertensive mice and found disease-associated, direction-dependent differences in directional complexity.  Similarly, Takahashi et al. \cite{Takahashi22} used persistent homology methods to analyze 3D structural images of murine lymphatic networks. Persistent homology has also been used to compute radiomic features in humans with chronic obstructive pulmonary disease (COPD)-affected airways \cite{Brodzki18}.

SSA includes methods that analyze the shape of objects independent of location, rotation, and scale. Most SSA methods use a collection of landmarks, a set of common points, to orient each object and then apply common statistical models to identify significant shape differences across the dataset. SSA has become increasingly popular for analysis of medical imaging data. Bruse et al. \cite{Bruse16} used an automated landmarking method to extract 3D anatomical shape features from cardiovascular magnetic resonance images of human aortas. In arterial networks, SSA has been used to study the right ventricular outflow tract \citep{Swanson2024}, lung morphology for aerosol inhalation dosimetry \citep{Xi2020}, brain arterial networks \citep{Bal2024, Guo2022}, and mean pulmonary arterial pressure (mPAP) \citep{Sabry23}. The last method is most similar to our application, since we apply SSA to distinguish healthy from hypertensive pulmonary arterial networks.

We explore the potential to diagnose pulmonary hypertension with each of these feature groups. We use ANCOVA tests for the geometric features, permutation tests for the topological data curves, and SVM with shape PCA scores for the SSA. We also explore the features' combined strength to diagnose pulmonary hypertension when used in a predictive model. We first put each of the computed features (and specifically chosen summaries of the TDA curves) into a sparse linear regression model and a boosted decision tree to identify the posterior inclusion probability and importance scores, respectively, for each variable. Then, we put the three most significant variables in each feature group into an SVM model, which is evaluated with leave-one-out cross-validation (LOOCV).

\section{Methods}

\begin{figure*}
    \centering
    \includegraphics[width=\textwidth]{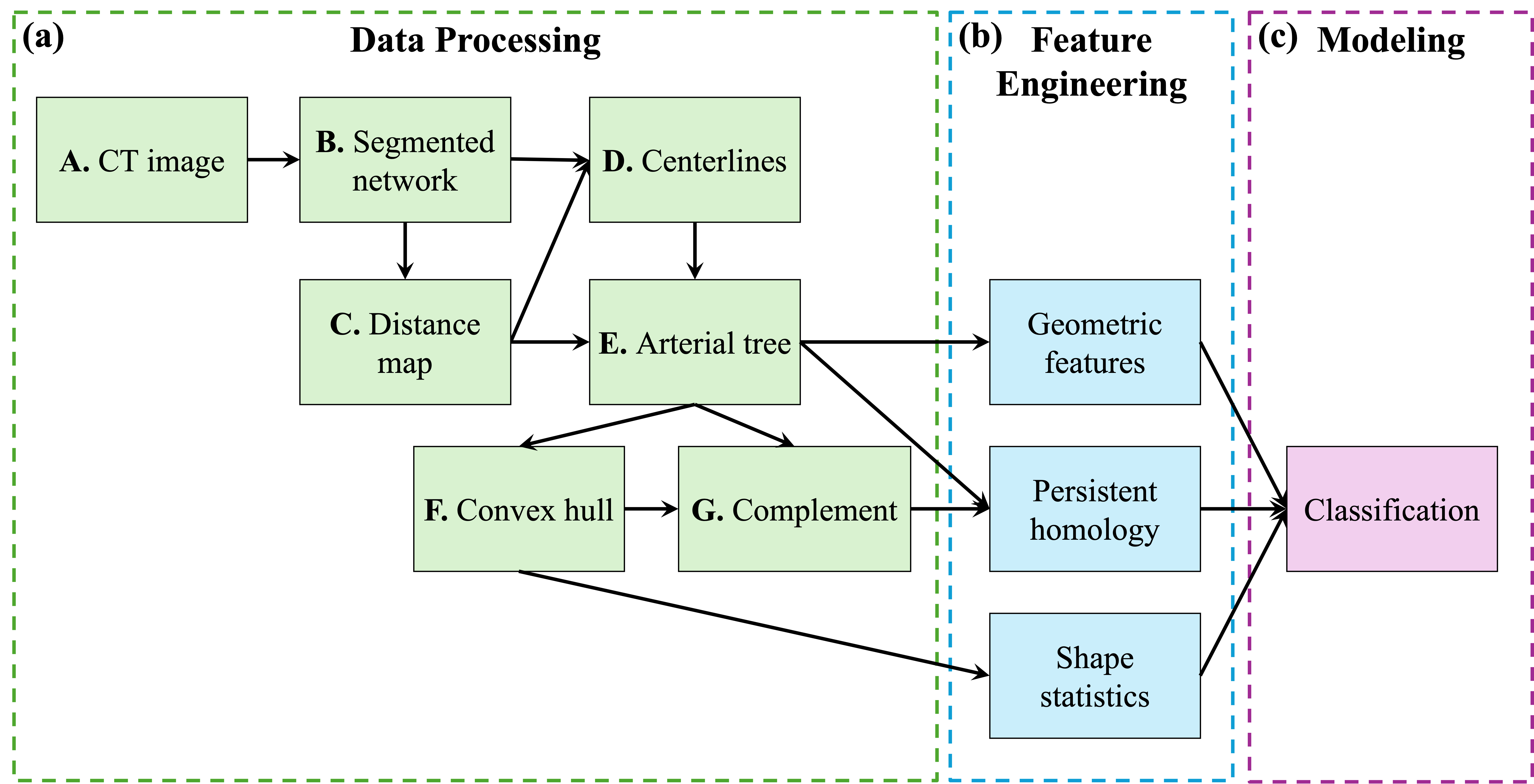}
    \caption{Feature extraction pipeline, including 3 phases. (a) Data processing: Micro-CT images (A) from control (CTRL) and hypertensive (HPH) mice are acquired and segmented, generating a 3D representation (B) of the arterial tree and its volume. Using skeletonization, we generate a distance map (C) and a centerline (D) through each vessel. An arterial tree (E) is generated, labeling each branch with its radius and length. We use a convex hull (F) to represent the volume the network occupies, and generate a complement (G) by subtracting the vessels from the convex hull.  (b) Feature engineering: Geometric, persistent homology, and SSA techniques are used to extract features describing the arterial network and lung volume. (c) Classification: Supervised machine learning is applied to phenotype control and hypertensive networks. Arrows indicate how structures and information are passed between tasks.}
    \label{fig:data-process}
\end{figure*}

\noindent This study focuses on extracting features from pulmonary arterial networks and the volume (the lung) spanned by these networks. We test the methodology by analyzing 3D renderings of the network and the lung extracted from micro-computed tomography (micro-CT) images. To test if the methodology can detect differences, we extract features from two groups of mice, comprising eight healthy controls (CTRL) and five mice with hypoxia-induced pulmonary hypertension (HPH).  
Feature extraction and analysis include: (i) generating arterial trees (labeled directional spatial graphs) and convex hulls from 3D renderings of the CT images, (ii) extracting geometric features, (iii) persistent homology and directional complexity, (iv) SSA, and (v) classification. These steps are outlined in Figure~\ref{fig:data-process} and are described in detail below. 

\subsection{Data Processing}

\begin{figure*}[h!tbp]
    \centering
        \includegraphics[width=0.85\textwidth]{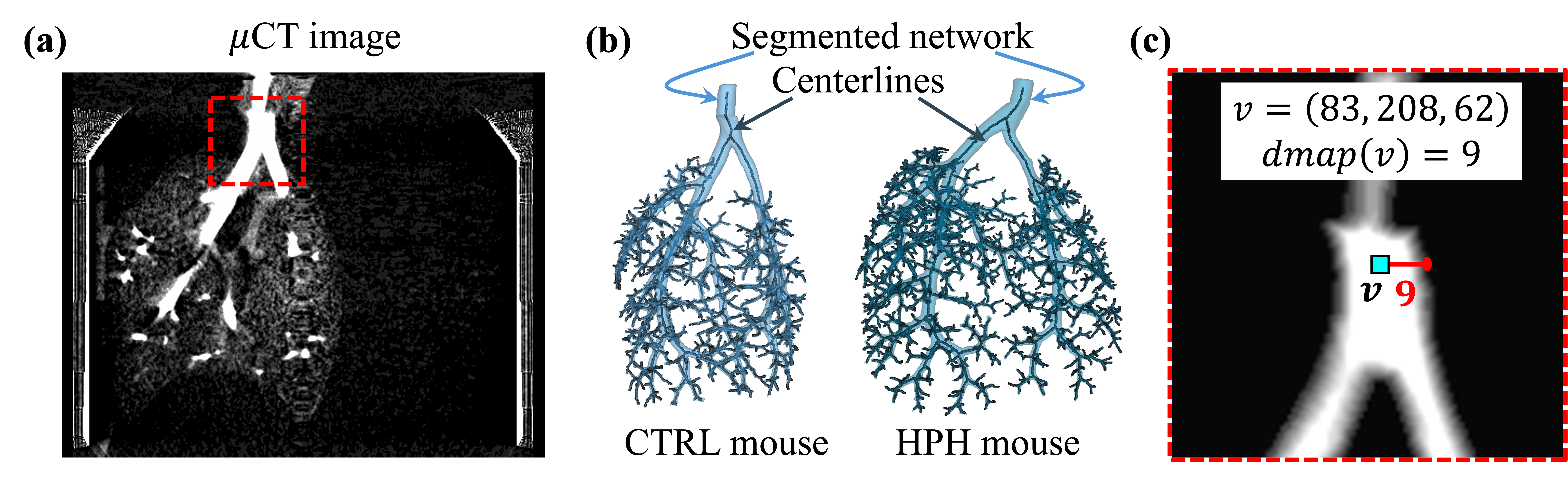}
    \caption{(a) Micro-CT image of a control mouse's pulmonary arteries. The figure shows the coronal view of the network, imaged at 6.3 mmHg. (b) 3D rendering (blue) and the vessel centerlines (black) of the arteries from the control (CTRL) mouse (from (a)) and a hypertensive (HPH) mouse. (c) Distance map (coronal view) zoomed in on the section marked with a red box in (a). The labeled voxel (cyan square) $v$ at the indicated $(x,y,z)$ coordinates is $9$ voxels away from the nearest background voxel. Hence $\text{dmap}(v)=9$. The corresponding distance in $\mu$m is obtained by multiplying $\text{dmap}(v)$ by the voxel-to-diameter scale obtained from the cannula inside the network.}
    \label{fig:ct-seg}
\end{figure*}

\smallskip
\noindent \textbf{A. Micro-CT Images.} The micro-CT images (example in Figure~\ref{fig:ct-seg}(a)) of the pulmonary arteries are acquired from eight control and five hypertensive mice. The images (made available by Naomi Chesler, University of California, Irvine) are obtained from male C57BL6/J mice aged 10-12 weeks with similar body weight. Pulmonary hypertension is induced in five mice by keeping them in a low-oxygen environment (50\% oxygen partial pressure) for 10 days. Details of the experimental protocol are provided in Vanderpool et al. \cite{Vanderpool11}. 

Before imaging, the mice are anesthetized and euthanized; the heart is removed, and the lungs are isolated and cleaned. The main pulmonary artery (MPA) is cannulated well above the first arterial bifurcation and perfused with perfluorooctyl bromide (PFOB) contrast. Each network is imaged via micro-CT, creating a 3D representation saved as a Digital Imaging and Communications in Medicine (DICOM) file. Images are obtained at four PFOB contrast pressures: 17.2 mmHg, 13.0 mmHg, 7.4 mmHg, and 6.3 mmHg. Each DICOM image has a resolution of 30-40 $\mu$m and comprises cubes of imaging data called voxels (analogous to pixels in a 2D image). Each voxel is located at a specific $(x,y,z)$ coordinate and has an intensity ranging from 0 (black voxel) to 255 (white voxel). 

\medskip
\noindent \textbf{B. Segmented Networks.} The micro-CT images represented by voxels (3D pixels) are segmented using 3D-Slicer \cite{Slicer, Fedorov12}. Segmentation involves differentiating foreground voxels (the white parts of Figure~\ref{fig:ct-seg}(a)) and background voxels (the black parts of Figure~\ref{fig:ct-seg}(a)). We identify foreground voxels using global thresholding. For this study, the threshold range was set to 25-255; all voxels with intensities in that range are included in the image foreground. All other voxels are assigned to the background. 

After global thresholding, the foreground voxels include the pulmonary arteries and non-biological features such as the imaging chamber (the white boundary shown in Figure~\ref{fig:ct-seg}(a)), glare, and noise. We remove the unwanted features and isolate the pulmonary arteries using a combination of cutting, erasing, and island-keeping. We do this minimally and carefully to avoid distorting the arterial structure. Lastly, we apply median smoothing with a kernel size of $3\times3\times3$ voxels. The segmented 3D representation is saved as a ``nearly raw raster data'' (NRRD) file containing foreground and background voxels. When visualized, it renders the foreground voxels in 3D, capturing all arteries that the PFOB contrast can reach.  Example 3D renderings of a control (CTRL) mouse and a hypertensive (HPH) mouse are shown in Figure~\ref{fig:ct-seg}(b).  

\medskip
\noindent \textbf{C. Distance Map. } Using the \texttt{dmap} function within the Spatial Graph Extractor package (SGEXT) \cite{SegExt}, we obtain a distance map for each segmentation. The distance map is saved as an NRRD file wherein each voxel is labeled with the distance from that voxel to the nearest background voxel (Figure \ref{fig:ct-seg}(c)). For each voxel $v$ in the segmentation, the distance map is given as
\begin{equation}
    \text{dmap}(v)=\min_{u\in\text{background}} ||u-v||_2.
    \label{eq:dmap}
\end{equation}
The distance map serves two purposes: it ensures the centerlines (see next section) are centered within each vessel and provides a radius measurement at each voxel along the centerline. 

\medskip
\noindent \textbf{D. Centerlines.} For each vessel in the segmented foreground, we generate a centerline (a collection of voxels) via skeletonization. We use the function \texttt{thin} (within SGEXT), a critical kernel method \cite{Couprie16}, to systematically remove voxels from each vessel until all vessels are one voxel wide. When faced with a choice of which voxel to remove, the algorithm uses the distance map to retain the voxel with the maximum dmap$(v)$ value (Equation \ref{eq:dmap}). This ensures that the skeletonized network (Figure~\ref{fig:ct-seg}b) is centered in the segmentation. The skeletonization, represented by the location of the voxels along the centerlines, is saved as an NRRD file.

\begin{figure}
    \centering
    \includegraphics[width=0.85\linewidth]{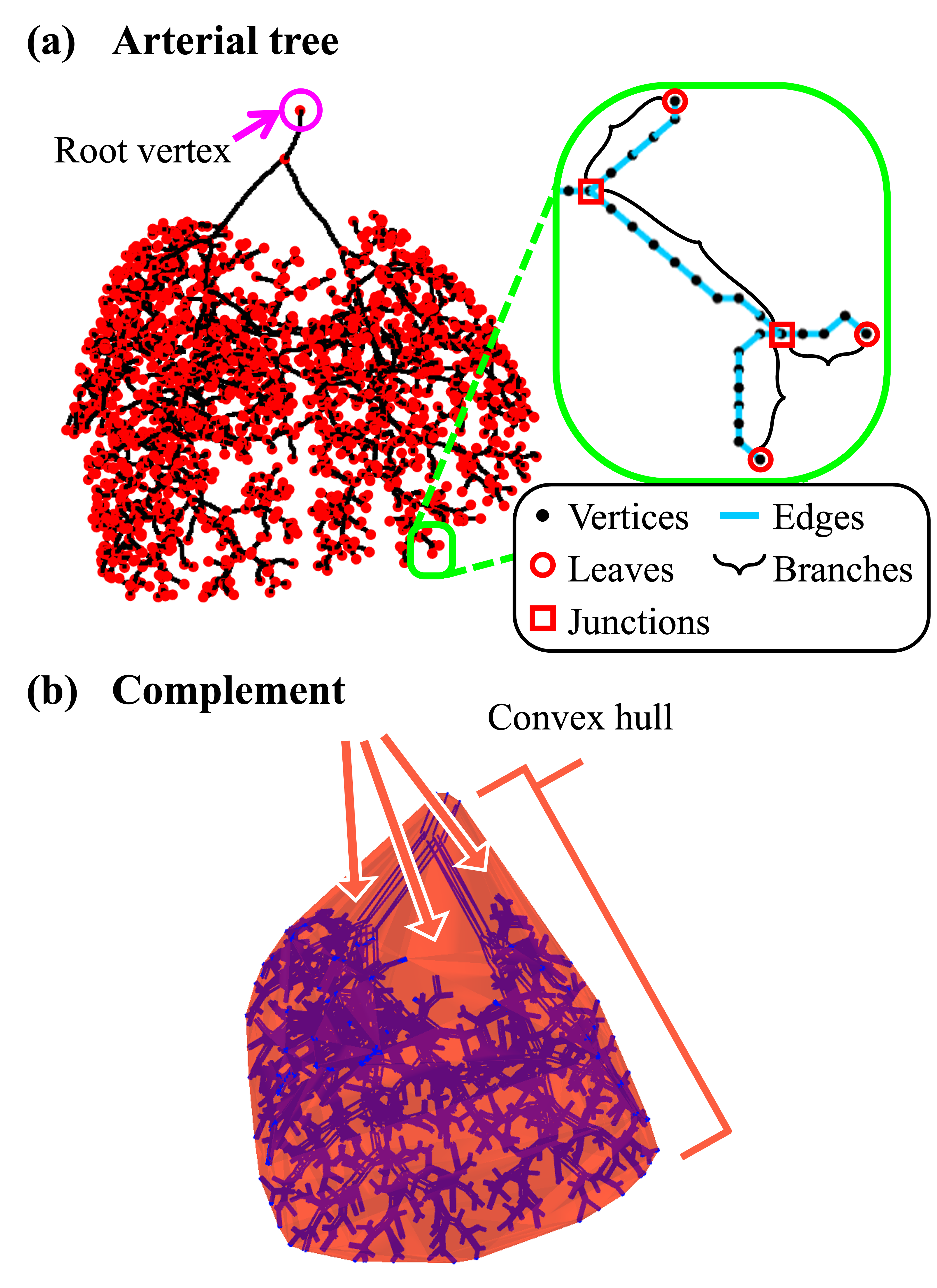}
    \caption{(a) Arterial tree from a hypertensive mouse at a pressure of 6.3 mmHg. A pink circle marks the root vertex. The green circle highlights four branches. Each branch (marked with a bracket) consists of vertices (black dots) connected by edges (blue lines). Vertices marked with red squares are junctions, and those marked with red circles are leaves. (b) Illustration of the convex hull (red volume) and complement (inside the red volume but outside the arterial tree (blue)) for the same hypertensive mouse in (a).}
\label{fig:tree-hull}
\end{figure}

\medskip
\noindent \textbf{E. Arterial Trees.} The SGEXT function \texttt{graph-extractor} uses the centerlines and the distance map to generate a directed spatial graph in $\mathbb{R}^3$ referred to as the \textit{arterial tree} (see Figure~\ref{fig:tree-hull}(a)). The tree is composed of a set of \textit{branches} (arteries), with a \textit{connectivity matrix} specifying the daughters (downstream branches) associated with each branch. Branches are composed of \textit{edges} connecting adjacent \textit{vertices} (the voxels along the centerline); each vertex exists at coordinates $(x,y,z)\in\mathbb{R}^3$. When combined in a tree, there are four vertex types: 
\begin{itemize}
\item  \textit{Root vertex.} Vertex of degree 1; the first vertex in the network representing the start of the main pulmonary artery (MPA).
\item \textit{Leaves.} Degree 1 vertices; each terminal branch ends with a leaf vertex.
\item \textit{Junctions:} Vertices of degree 3 or greater. These vertices are found at the $(x,y,z)$ coordinates where three or more branches meet.
\item \textit{Internal vertices:} Degree 2 vertices; forming all branches in the network. For each branch these are the 2nd to 2nd to last vertex along each branch. 

"Internal vertices: Degree-two vertices that form the interior of each branch"

\end{itemize}

The first branch in the connectivity matrix represents the MPA, which starts at the root vertex. All branches are oriented away from the root in the direction of blood flow; thus, each branch has a top and bottom vertex. For the arterial trees generated in this study, a branch can have up to 4 daughters, though the vast majority of branches ($> 95\%$) have 2 daughters. Junctions with 2 daughters are referred to as \textit{bifurcations}, those with 3 daughters as \textit{trifurcations}, and those with 4 daughters as \textit{quadfurcations}. 

The arterial tree is post-processed to remove errors \cite{Chambers20}, including small cycles of nearby vertices, false edges, and blobs (collections of short branches created around a junction vertex). The removal of small cycles and false edges is described in \cite{Chambers20}. Blobs are removed by identifying a vertex upstream, downstream, and in the middle of the blob. Similar to the removal of small cycles \cite{Chambers20}, blob removal uses Dijkstra's shortest path algorithm \cite{Dijkstra59} to identify the vertices and edges along the shortest path between the input and output vertices. We remove every vertex and edge not included on the shortest path within 5 voxels ($\sim$ 200 $\mu$m) of the blob center. 

Each vertex in the tree is labeled with the vessel radius at that point (the \texttt{dmap} value of the corresponding voxel, converted to $\mu $m). Each branch is labeled with its total length, computed as the sum of the lengths of all its edges. Hence, the arterial vessels, represented by a large number of voxels in the 3D rendering, are reduced to a branch in a graph with a radius at each vertex. In the following, we use \textit{branch} when we refer to the collection of edges in the graph and \textit{vessel} when we refer to the object in the 3D rendered representation.

\medskip
\noindent\emph{Branch radius.} Each branch is assigned a single radius, obtained using the statistical change point algorithm developed by Bartolo et al. \cite{Bartolo2024}. For each branch, this algorithm reads the vector of $n$ vertices and their corresponding radii. The radius vector is fed into a script, which uses two standard R packages, \texttt{changepoint} and \texttt{segmented} 
\cite{killick2014changepoint,muggeo2008segmented}, to determine and place up to three change points along the branch. The algorithm identifies the slope of each line segment generated by a linear fit between the vessel’s change points and its endpoints. The algorithm identifies the segment with the smallest slope, and the branch's radius and standard deviation are set to the arithmetic mean and standard deviation of the radius values within that segment. 

\begin{figure*}[h!tbp]
\centering
    \includegraphics[width=0.85\textwidth]{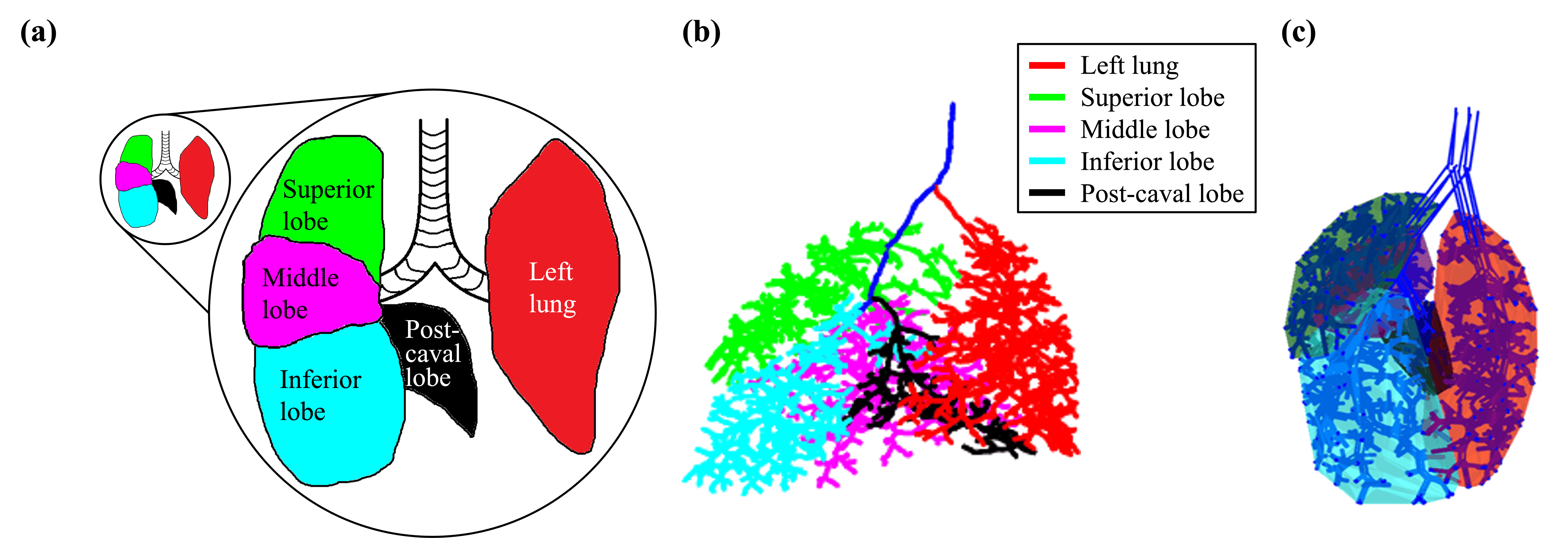}
\caption{(a) Mouse lung anatomy. The left lung has one lobe; the right lung has four lobes: superior, middle, inferior, and post-caval. (b) Example network from a control mouse with branches color-coded to show which lobe they belong to. (c) Example convex hulls of each lobe, color-coded accordingly.}
    \label{fig:lobes}
\end{figure*}

\medskip
\noindent\emph{Lobe trees.} As shown in Figure~\ref{fig:lobes}(a), the mouse lung has five lobes: left, superior, inferior, middle, and post-caval. Using visual inspection, we identify the starting branch for each lobe. The lobe tree contains all connected branches downstream from the starting branch. Figure~\ref{fig:lobes}(b) shows the trees associated with each lobe. 

\medskip
\noindent\emph{Pruning. }  Similar to our previous study \cite{Miller23}, we employ pruning when comparing the arterial trees obtained from the control and hypertensive animals. Because the hypertensive animals have larger radii \cite{Chambers20}, the PFOB contrast perfuses further down the tree, making more vessels visible in the micro-CT image. As a result, the arterial trees representing the hypertensive animals contain more branches than the ones constructed from the controls. To normalize the networks, for each pressure, we remove pairs of terminal branches whose radii are below a specified threshold. The threshold is iteratively increased, and the process is continued until the number of branches in the pruned hypertensive tree equals the average of the corresponding control trees. Radii values for the pruning of each hypertensive mouse can be found in Table S1 in the Supplementary Materials.

\medskip
\noindent \textbf{F. Convex Hull.} In the micro-CT images analyzed here, the vasculature is separated from the lungs before imaging \cite{Vanderpool11}; thus, we cannot generate a 3D rendering of the lobe or lung volume. Therefore, to approximate the lung and lobe volumes, we compute the convex hull of the vertices within each lobe (shown in Figure~\ref{fig:lobes}(c)), the arterial tree (shown in Figure~\ref{fig:tree-hull}(b)), and the lobe trees (shown in Figure~\ref{fig:lobes}(c)). The convex hulls are the smallest convex volume containing the vertices in question. They are created by inputting the voxel clouds into SciPy's ConvexHull function \cite{Scipy}. 

\medskip
\noindent\textbf{G. Complement.} Remodeling induced by pulmonary hypertension begins in the distal pulmonary circulation \cite{Tobal1993}, i.e., the small pulmonary arteries, arterioles, and precapillary vessels are exposed to remodeling before the larger arteries. These smallest vessels belong to spaces not visible in micro-CT images. Following the terminology proposed by Belchi et al.~\cite{Brodzki18}, we refer to this space between the arterial tree and the enclosing lung boundary as the lung/lobe complement. Since anatomical lobe surfaces are unavailable in our excised pulmonary arterial data, we approximate the enclosing lobe and lung boundary by the convex hull of the lobe or lung specific arterial point cloud, i.e.,  the ``lung/lobe complement'' point cloud consists of:  convex-hull boundary points, and five arterial-wall points sampled on the circumference of each centerline point. For each branch, we obtain points along the vessel boundary by sampling the vertices at 20\,\(\mu\)m intervals and choosing points according to the average radii along the vessel. 

Geometrically, this point cloud contains points on the arterial surface and the boundary of the convex-hull surrogate for the lung or lobe boundary. Its persistent homology therefore provides a proxy for the organization of the unresolved region, which may contain distal vessels not visible in the micro-CT images.  

\subsection{Geometric Features} \label{Msec:GeomFeat}
\smallskip 

\noindent Since Murray's seminal work \cite{Murray26}, researchers have studied the morphometry of arterial trees extensively using geometric descriptors of vascular structure. Most of these are derived based on the assumption that vascular development reflects underlying optimization principles. In the pulmonary circulation, these principles include minimizing the work required to distribute blood \cite{Murray26}, minimizing total lumen volume or surface area, and minimizing pumping power or drag force \cite{Zamir76}.

\begin{figure}
    \centering
        \includegraphics[width=1\linewidth]{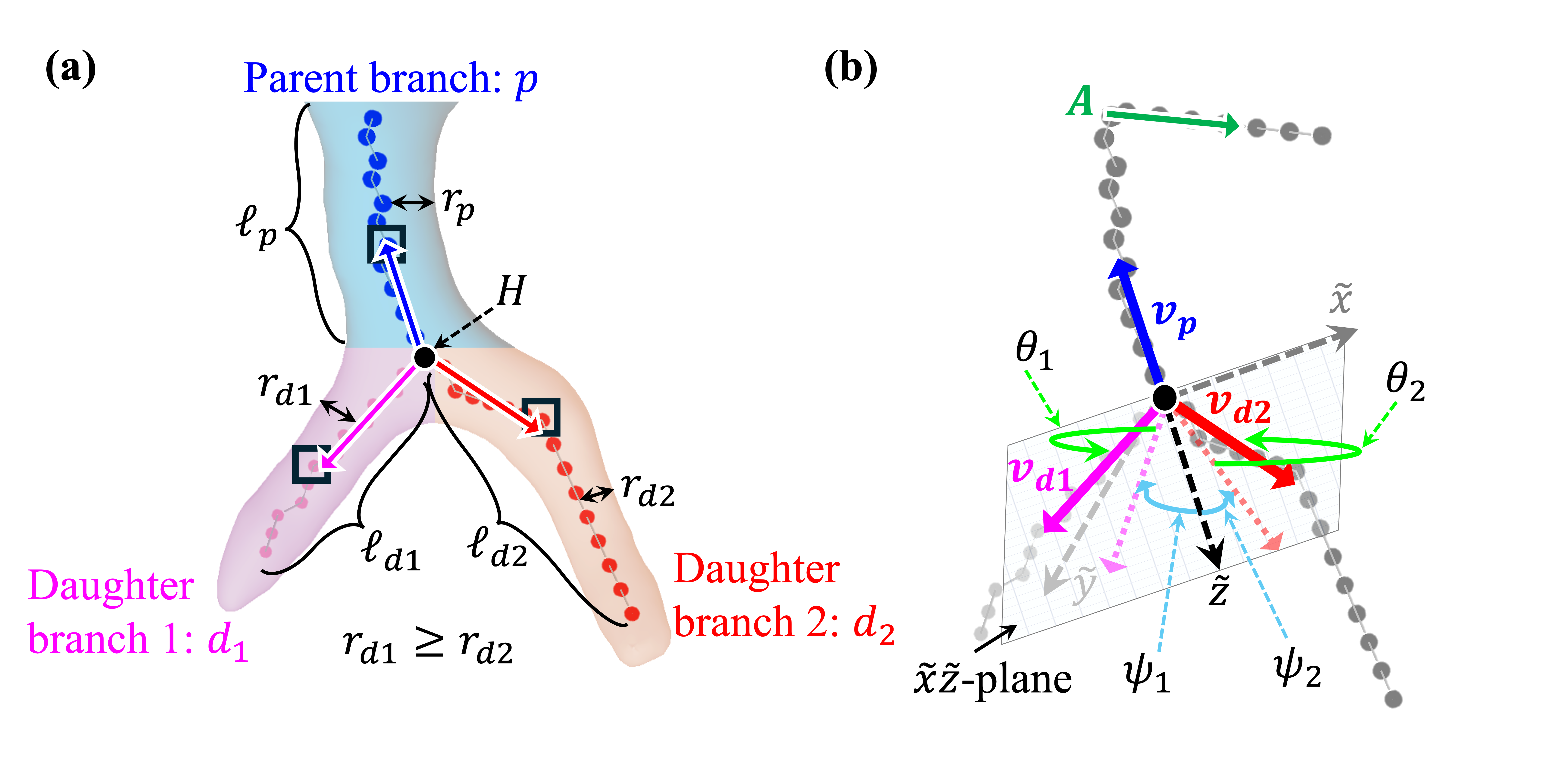}
    \caption{(a) Example bifurcation, showing the segmented vessels and centerlines. The radii $r$ and lengths $\ell$ of the parent (subscript $p$) and daughter (subscript $d1,d2$) vessels are labeled. Daughter 2 is always the daughter with the smaller radius. $H$ denotes the hub node of this bifurcation. Black squares mark the change points, and the vectors they create are marked with blue (parent vessel vector), pink (daughter 1), and red (daughter 2) arrows. (b) Branching angles: $\psi_{1,2}$ denote the in-plane angles between $\tilde{z}$ (the reverse direction of $\mathbf{v_p}$) and the projections of $\mathbf{v_{d1,d2}}$ in the $\tilde{x}\tilde{z}$-plane (which is also the parent-aunt plane); $\theta_{1,2}$ denote the out-of-plane angles needed to rotate those projections to $\mathbf{v_{d1,d2}}$.}
    \label{fig:geometry}
\end{figure}

While these principles hold for the healthy lung, they likely deviate for the diseased lung due to vascular remodeling. To uncover the details of this remodeling, we examine geometric features in control and hypertensive arteries, analyzing the arterial tree, convex hulls, and lung complements. The majority of the geometric features (summarized in Table \ref{table:geometric_features}) are calculated under the assumption that all junctions are bifurcations (i.e., one ``parent'' branch splits into two ``daughter'' branches; see Figure~\ref{fig:geometry}). For the trees analyzed here, as reported in our previous study \cite{Chambers20}, this assumption is satisfied for more than 95\% of the junctions \cite{Chambers20}. For features that assume a bifurcating structure, we ignore tri- and quad-furcations and label branches as shown in Figure~\ref{fig:geometry}. 

At each bifurcation, the radius of the parent branch is denoted by $r_p$ and the radii of its two daughter branches are denoted by $r_{d1}$ and $r_{d2}$. The branches are labeled such that daughter 2 has a radius less than or equal to daughter 1. Similarly, the lengths are denoted by $\ell_p$, $\ell_{d1}$, and $\ell_{d2}$.  The junction hub node is denoted by $H$, and vectors defining the direction of the branches are denoted $(\bf{v}_p,\bf{v}_{d1},\bf{v}_{d2})$. To capture the vessel vectors' directions relative to the bifurcation at $H$, we insert up to one change point along each vessel at the vertex that minimizes the least-squares error between the lines of best fit through the branch points. The vectors $\mathbf{v_i}$  ($i=p, d1, d2$) are taken to be in the direction of the line of best fit (found using linear regression) for the edge points between $H$ and the change point. In Figure~\ref{fig:geometry}(a), squares mark the change points, and the $\mathbf{v_i}$ vectors, connecting $H$ to the change points, are shown with arrows. 

\begin{table*}[bp]
\centering
\caption{List of features organized by type: geometric, topological, and shape. The last column represents the level at which the computation is performed: V - branch, B - bifurcation, L/M denotes a particular lobe or the whole mouse arterial network, D -dataset.}
\renewcommand{\arraystretch}{1.1}
\begin{tabular}{ c |  c  c  c }
\toprule
\textbf{} &
\textbf{Feature} &
\textbf{Feature Description} &
\textbf{Computation} \\ \midrule
\multirow{14}{*}{\centering\makebox[0pt][c]{\rotatebox[origin=c]{90}{Geometric}}}
& r [$\mu$m] & Radius & V \\
& $\ell$ [$\mu$m] & Length & V \\
& V [$\mu\text{m}^3$] & Vessel volume (Eq.\ref{eq:volume}) & V \\
& $\alpha$ & Radius ratio (Eq. \ref{eq:alpha}) & B \\
& $\beta$ & Radius ratio (Eq. \ref{eq:alpha}) & B \\
& $\xi$ & Radius exponent (Eq. \ref{eq:xi}) & B \\
& $\gamma$ & Asymmetry ratio (Eq. \ref{eq:gamma}) & B \\
& $\eta$ & Area ratio (Eq. \ref{eq:eta}) & B \\
& $\psi_{i=1,2}$ [rad] & In-plane angle (Eq. \ref{eq:psi}) & B \\
& $\theta_{i=1,2}$ [rad] & Out-of-plane angle (Eq. \ref{eq:theta}) & B \\
& D & Tree Depth & V \\
& $L_R$ [$\mu$m] & Length-from-root & V \\
& $\kappa$ & Tortuosity (Eq. \ref{eq:tort}) & V \\
& $\ell_{rr}$ & Length-to-radius ratio (Eq. \ref{eq:lrr}) & V \\
\hline
\multirow{4}{*}{\centering\makebox[0pt][c]{\rotatebox[origin=c]{90}{TDA}}}
& Landscape Curve & Pointwise \(k\)-th largest tent-function value & L/M \\
& Betti Curve & Number of generators & L/M \\
& Lifespan Curve & Persistence & L/M \\
& Norm Lifespan Curve & Normalized persistence & L/M \\ \hline
\multirow{5}{*}{\centering\makebox[0pt][c]{\rotatebox[origin=c]{90}{Shape}}}
& $\Tilde{{\bf P}}$ & Point Cloud & L/M \\
& ${\bf C}(\Tilde{{\bf P}})$ & Convex hull & L/M \\
& $\Tilde{{\bf X}}$ & L landmarks & L/M \\
& ${\bf Z}$ & $n\times J$ Summary of PC Scores & D \\
& ${\bf Z}_k^T$ & $k^{th}$ row of ${\bf Z}$ & L/M \\ \bottomrule
\end{tabular}
\label{table:geometric_features}
\end{table*}

\medskip
\noindent\textbf{Vessel Volume.} For each vessel, the volume ($V_j$, $\mu^3$) is computed as the sum of a series of conical segments along the vertices within the centerline. Each vertex with position ${\bf x}_i = (x_i,y_i,z_i)$ along the centerline is associated with a radius $r_i$ (mm) and a length $\ell_i$ denoting the Euclidean distance from the beginning of the vessel ${\bf x}_1$ to ${\bf x}_i$. For a vessel with $n$ vertices along the centerline, the total edge volume is computed as  
\begin{equation}
   V_j = \sum_{i=1}^{n-1} \frac{\pi}{3}
   \left(\ell_{i+1}-\ell_i\right)
   \left(r_{i+1}^2 + r_i^2 + r_{i+1} r_i\right).
   \label{eq:volume}
\end{equation}
This measure provides insights into spatial distribution and relative size within the network \cite{Mackenzie25}. Note that the vessel volume differs from the edge volume computed using the radius assigned to each vessel. 

\medskip
\noindent\textbf{Radius Scaling.} Murray \cite{Murray26, Murray26a} showed that the energy in a bifurcating network is minimized when the radius of the parent branch relative to its daughters is related by \mbox{$r_p^3 = r_{d1}^3 + r_{d2}^3$.} Uylings generalized this relation \cite{Uylings77} to 
\begin{equation}
   r_p^\xi = r_{d1}^\xi + r_{d2}^\xi, \ \ 1\leq\xi\leq 3. 
   \label{eq:xi}
\end{equation}
For arterial vascular networks, Uylings found that $\xi=2.76$. Other studies \cite{Murray26,Murray26a,Uylings77,William07,Correa25} found that $\xi$ ranges from $2$ to $3$. To estimate $\xi$, we rewrite Equation \ref{eq:xi} as
\begin{equation}
    0 = f(\xi) = \left(\frac{r_{d1}}{r_{p}}\right)^{\xi} + \left(\frac{r_{d2}}{r_{p}}\right)^{\xi} - 1.
\end{equation}
Using the radii assigned to each branch, we compute $\xi$ numerically using Newton's method \cite{Akram15} 
\begin{align*}
    \xi_{i+1} &= \xi_i - \frac{f(\xi_i)}{f'(\xi_i)} \\*
              &= \xi_i -
    \frac{\left(\frac{r_{d1}}{r_p}\right)^{\xi_i} + \left(\frac{r_{d2}}{r_p}\right)^{\xi_i} - 1}{\left(\frac{r_{d1}}{r_p}\right)^{\xi_i}\ln{\left(\frac{r_{d1}}{r_p}\right)} + \left(\frac{r_{d2}}{r_p}\right)^{\xi_i}\ln{\left(\frac{r_{d2}}{r_p}\right)}}.
\end{align*}
This method is iterated until a $\xi$ is found with an absolute residual satisfying the prescribed tolerance $|f(\xi)| \leq 10^{-5}$. 

This method is numerically unstable when the parent and daughter branches are of similar radii. Therefore,  $\xi$ is only calculated for bifurcations in which both daughter branches have radii smaller than $95\%$ of the parent radius.

\medskip
\noindent\textbf{Area and Asymmetry Ratios.} Zamir \cite{Zamir78} and Olufsen \cite{Olufsen98} define the asymmetry ratio as the smaller daughter's radius divided by the larger daughter's radius. Following these studies, we denote the asymmetry ratio as 
\begin{equation} 
   \gamma = \frac{r_{d2}^2}{r_{d1}^2}, \ \ 0 < \gamma \leq 1.
\label{eq:gamma}
\end{equation}
This factor captures bifurcation symmetry. In a bifurcation where the daughter vessels have equal radii, $\gamma=1$. In a bifurcation where one daughter vessel has a significantly higher radius than the other, $\gamma\approx0$. Zamir \cite{Zamir78} and Olufsen \cite{Olufsen98} further define the area ratio between the daughter vessels and the parent vessel as 
\begin{equation}
    \eta = \frac{r_{d1}^2+r_{d2}^2}{r_p^2}.
    \label{eq:eta}
\end{equation}
$\eta$ denotes how the area is conserved after a bifurcation. Values closer to $1$ imply better conservation. In arterial networks, the total area increases with the number of bifurcations; thus, it is assumed that $\eta>1$.

\medskip
\noindent\textbf{Radius Ratios.} A third expression describing asymmetry is via the radius scaling factors relating the radii of the parent and daughter vessels as
\begin{equation}
   r_{d1} = \alpha r_p, \ \ \  r_{d2} = \beta r_p. \label{eq:alpha}
\end{equation}
Using our assumption that $r_{d1} > r_{d2}$ gives that $\alpha \geq \beta$. As noted by Olufsen \cite{Olufsen00}, the five bifurcation scaling parameters: $(\xi, \eta, \gamma, \alpha, \beta)$  are not independent. For example, the radius scaling factors $\alpha$ and $\beta$ satisfy that 
\begin{equation}
  \alpha = \left(1+\gamma^{\xi/2}\right)^{-1/\xi}, \ \ \
  \beta  = \alpha \sqrt{\gamma}.
  \label{eq:alphabetaGamma}
\end{equation}
In addition to being determinable from $\gamma,\xi,$ and $\eta$, these factors can be determined directly from the data. 

\medskip
\noindent\textbf{Length-to-Radius Ratio.} Another geometric feature, calculated for each vessel, is the length-to-radius ratio 
\begin{equation}
 \ell_{rr} = \frac{\ell}{r},\label{eq:lrr}
\end{equation}
where $\ell$ and $r$ denote the length and radius of the vessel; for arteries we assume that $\ell_{rr} > 1$.

\medskip
\noindent\textbf{Tortuosity.} In the pulmonary vascular network, the blood vessels are curved. Tortuosity is a method of quantifying the curvature in a vessel \cite{Helmberger2014}. It can be calculated as
\begin{equation}\label{eq:tort}
\kappa = \frac{\ell}{\ell^*}, 
\end{equation}
where $\ell$ denotes the Euclidean distance along the vessel, and 
\[
  \ell^* = \sqrt{((x_n-x_1)^2 + (y_n-y_1)^2 +(z_n-z_1)^2)}
\]
denotes the length of the straight vessel spanned between the first $(x_1,y_1,z_1)$ and last $(x_n,y_n,z_n)$ voxel within each vessel. A value of 1 denotes a vessel with no curvature, and larger values indicate greater curvature. This measure may be biased for very short vessels.

\medskip
\noindent\textbf{Branching Angles.} Branching angles can be used to characterize how the three-dimensional structure of the arterial tree changes with disease. At each bifurcation, we use the vectors $(\bf{v}_p,\bf{v}_{d1},\bf{v}_{d2})$ to calculate branching angles $\psi_{1,2}$ and $\theta_{1,2}$ using the methods in \cite{Lee18}. We also define a vector $\mathbf{v_A}$, representing the direction of the ``aunt vessel''. This vessel is the accompanying daughter vessel at the previous bifurcation (the parent's sibling). To do this, we reorient the frame of reference (standard $x$-,$y$-,$z$-axes) to an adjusted $\tilde{x}$-$\tilde{y}$-$\tilde{z}$ reference frame where 
\begin{equation}
  \tilde{z} = -\frac{\mathbf{v_p}}{\|\mathbf{v_p}\|}, \ \ \ 
  \tilde{x} = \frac{\mathbf{v_A}-(\mathbf{v_A}\cdot \tilde{z})\tilde{z}}{\|\mathbf{v_A}-(\mathbf{v_A}\cdot \tilde{z})\tilde{z}\|}, \ \ \
  \tilde{y} = \tilde{z}\times\tilde{x}.
\end{equation}
The above equations can be obtained by reorienting the coordinate system such that the $\tilde{z}$-axis is the unit vector pointing in the direction of the parent vessel (away from the bifurcation) and the $\tilde{x}$-axis is the unit vector lying in the parent-aunt plane, the plane passing through the bifurcation node $H$ and vectors $\mathbf{v_p}$ and $\mathbf{v_A}$.

In the new coordinate system, each daughter vessel $i=1,2$ at a bifurcation can be characterized by an in-plane angle $\psi_i$ and an out-of-plane angle $\theta_i$. With our adjusted axes, the $\psi_i$ angles are the angles between the daughter vessels and $\tilde{z}$ as defined within the parent-aunt plane. We calculate them using the following equation \cite{Lee18}
\begin{equation}
    \psi_i = \cos^{-1}\left(\frac{\tilde{z}\cdot\mathbf{v_i}}{\|\tilde{z}\|\|\mathbf{v_i}\|}\right).\label{eq:psi}
\end{equation}
The out-of-plane angles, $\theta_{1,2}$, are similarly defined as the clockwise angles of rotation from the $\tilde{x}\tilde{z}$-plane around the $\tilde{z}$ axis. We calculate this angle by
\begin{equation}
    \theta_i = \pi -\tan^{-1}\left(\frac{\mathbf{v_i}\cdot\tilde{y}}{\mathbf{v_i}\cdot{\tilde{x}}}\right).\label{eq:theta}
\end{equation}

\medskip
\noindent\textbf{Tree Depth.} The depth of a vessel \cite{bollobas1998modern} is defined as the number of bifurcations along the unique path from the root to the vessel. This statistic enables analysis of vessel distributions as a function of hierarchical position within the network. Vessel depths are computed using a breadth-first search (BFS) traversal of the tree \cite{cormen2022introduction}. The traversal starts at the root vessel at depth 0, and each daughter vessel is assigned a depth one greater than its parent, ensuring that all vessels at depth $k$ are processed before vessels at depth $k+1$. For lobe-specific analyses, the depth of a vessel is measured relative to the corresponding lobe input vessel.

\medskip
\noindent\textbf{Length-from-root.} The length-from-root statistic measures the distance of a vessel’s inlet from the origin of the vascular network. This is computed by summing the lengths of the vessels along the shortest path from the origin to the given vessel. This measure provides an alternative perspective on how vessel radius may decrease, focusing on the physical distance from the origin rather than depth alone. Depth treats all vessels at the same hierarchical level (i.e., vessels with the same depth) as being equidistant from the origin, which this metric aims to overcome. When calculated at the lobe level, this metric measures the distance from the input vessel to the lobe.

\medskip
\noindent\textbf{Classification with Geometric Features.}  To evaluate whether any of the geometric features alone can discriminate between control and pulmonary hypertension, we formulate the following statistical model. For networks with pressure $i$ from mouse $j$ in treatment group $k$, the feature $Y_{ijk}$ is determined by
\begin{equation*}
  Y_{ijk} = [b_0 + \tau_{0k} + U_{0j}] + [b_1 + \tau_{1k} + U_{1j}]P_i  + E_{ijk}, 
\end{equation*}
with normally distributed random effects
\[
\quad E_{ijk} \sim \mathcal{N}(0, \sigma^2_E)\qquad
\begin{bmatrix}
U_{0j} \\
U_{1j} 
\end{bmatrix} \sim \mathcal{N}\left({\bf 0}, {\bf \Sigma}\right)
\]
at levels of pressure:
\[
\text{Pressure}_i = \begin{cases}
6.3 & i = 1 \\
7.4 & i = 2 \\
13.0 & i = 3 \\
17.2 & i = 4. 
\end{cases}
\]
Here, $b_0$ and $b_1$ denote the intercept and slope of the features with respect to the pressure (in mmHg). The parameters $\tau_{0k}$ and $\tau_{1k}$ represent the intercept and slope of the $k^{th}$ treatment group (hypertensive vs. control), and the random effects $U_{0j}$ and $U_{1j}$ are the mouse-specific random effects for the intercept and slope associated with mouse $j$. 

To test whether the median and standard deviation of the features differ significantly between the hypertensive and control groups, we perform $t$-tests to check whether $\tau_{0k}$ and $\tau_{1k}$ differ from zero. The $t$-test statistics are computed with
\[
  \frac{\hat\tau }{SE(\hat\tau)}
\]
and the p-values are the probabilities of obtaining a t-statistic as extreme as or more extreme than the observed value. This procedure is performed for the entire network from each mouse and for the five separate lobes for each geometric feature.  

\subsection{Persistent Homology}\label{Msec:PH}
\smallskip

\noindent Persistent homology characterizes how topological features of data, including connected components, cycles, and cavities, appear and disappear as a filtration parameter varies \cite{nanda2013simplicial}. Each feature is represented by a persistence interval that records the birth and death filtration values. Their persistence or lifespan is the length of this interval. The resulting collection of intervals can be represented by a barcode or persistence diagram and summarized using quantities such as total persistence, Betti curves, lifespan curves, and persistence landscapes.

\begin{figure*}
\centering
\includegraphics[width=0.9\linewidth]{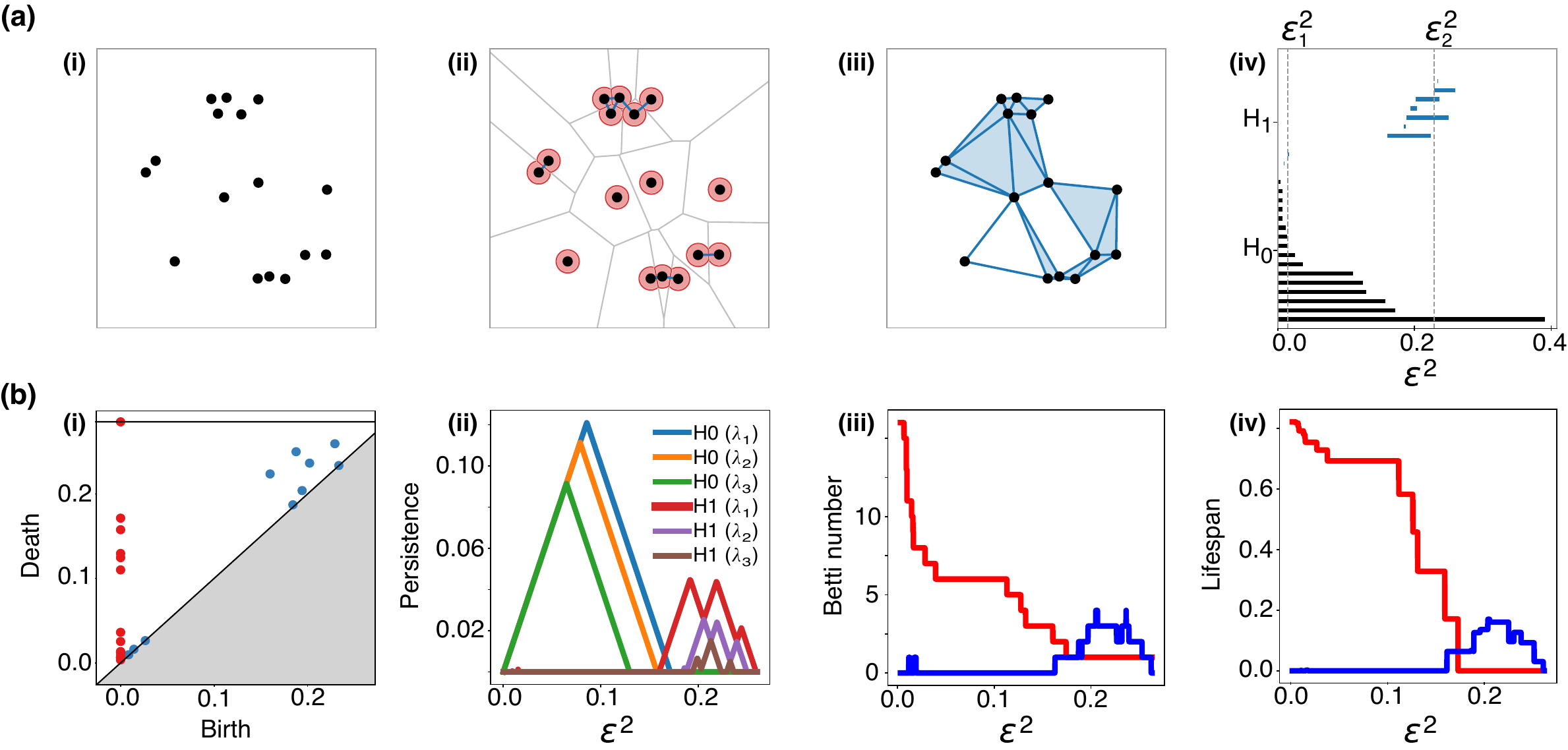}
\caption{(a) Summary of the alpha-complex filtration. (i) Sample point cloud. (ii) Voronoi cells and $\varepsilon$-balls at $\varepsilon_1=0.118$. (iii) Alpha complex at $\varepsilon_2=0.478$. (iv) The resulting persistence barcodes in homological dimensions $0$ and $1$. (b) (i) Example persistence diagram created from the points in Figure~\ref{fig:PH}(a) with birth-death pairs in dimension 0 (red) and 1 (blue). (ii) Persistence Landscape with the first three landscapes in dimensions 0 and 1. (iii)-(iv) Betti and Lifespan curves, with the dimension 0 curve in red and dimension 1 in blue.}
\label{fig:PH}
\end{figure*}

In this study, we use alpha-complex filtration to analyze the arterial and complement point clouds in homological dimensions $0$, $1$, and $2$. As the scale parameter $\epsilon$ (mm) increases, the alpha complex captures the point cloud's evolving topology while respecting its underlying geometry; see Figure~\ref{fig:PH}(a). Formal definitions of the alpha complex, persistent homology groups, persistence intervals, and total persistence are provided in the Supplementary Materials Section 1.1.

Although the underlying arterial graph is a tree, its alpha-complex filtration may have nontrivial $H_1$ and $H_2$. These features arise from the spatial embedding of the centerlines as nearby branches interact with increasing filtration scale, rather than from cycles in the arterial graph itself.

\medskip
\noindent\textbf{Persistence Diagrams.}  Rather than listing all persistent homology groups individually, the evolution of homological features can be summarized using a persistence diagram or a barcode. The $p$th persistence diagram associated with the filtration induced by $f$, denoted $\operatorname{Dgm}p(f)$, is the multiset consisting of all pairs $(a_i,a_j)$, where a $p$-dimensional homology class is born in $X_i$ and dies upon entering $X_j$, together with the diagonal in $(\mathbb{R}{\ge 0}\cup{\pm\infty})^2$, see Figure~\ref{fig:PH}(b) i. Each point $(b,d)$ in the diagram is a \textbf{birth--death pair}, recording the birth time $b$ and death time $d$ of a persistent homology class. Barcodes provide an alternative way to present this information; see Figure~\ref{fig:PH}(a) iv. 

\medskip
\noindent\textbf{Persistence Curves.} A persistence diagram is a finite multiset of points whose cardinality may vary across observations, and therefore it does not naturally admit a fixed-dimensional vector-space representation. Consequently, standard statistical operations and many conventional machine learning methods cannot be applied to persistence diagrams without additional structure. Therefore, to facilitate statistical analysis, persistence diagrams are often transformed into representations that are compatible with standard statistical and machine learning methods. Several approaches have been proposed to do so, including persistence curves~\cite{chung2019persistence}, persistence kernels~\cite{reininghaus2015stable}, persistence landscapes~\cite{bubenik2015statistical}, and persistence images~\cite{adams2017persistence}. In this study, we focus on persistence curves, which provide functional summaries of persistent homology that can subsequently be analyzed using standard statistical techniques.  We compute Betti, lifespan, and normalized lifespan curves, and persistence landscapes.

Let $D_n$ denote the persistence diagram in homological dimension $n$. The $n$th \textbf{Betti curve}, denoted by $\beta_n(D_n,t)$, is determined by the number of birth-death pairs $(b,d)\in D_n$ assuming that $t\in(b,d]$ with $d<\infty$, see Figure~\ref{fig:PH}(b) iii. Similarly, the $n$th \textbf{lifespan curve}, denoted by $\ell_n(D_n,t)$, is the persistence-weighted Betti curve, defined as the sum of the lifespans $d-b$ of all finite birth-death pairs $(b,d)\in D_n$ for $t\in(b,d]$. Thus, whereas each active class contributes unit weight to the Betti curve, its contribution to the lifespan curve is weighted by its persistence, so that longer-lived topological features have greater influence; see Figure~\ref{fig:PH}(b) iv. The normalized lifespan curve is obtained by dividing the lifespan curve by the total persistence. For the point cloud representing the arterial tree, the dimension-one ($n=1$) curves capture branch-to-branch proximity \cite{Brodzki18}. For the arterial centerline point cloud, dimension-one persistence characterizes interactions between spatially nearby branches. Following \cite{Brodzki18}, we define branch-to-branch proximity as the total finite $H_1$ persistence,
    $\operatorname{B2B}(D_1)
    =\sum_{\substack{(b,d)\in D_1\\d<\infty}}(d-b).$
The corresponding $H_1$ persistence curves and landscapes provide functional summaries of how these features are distributed across filtration scales.

Finally, we utilize the definition of persistence landscapes introduced by Bubenik~\cite{bubenik2015statistical}. The $k$th \textbf{persistence landscape} associated with $D$, denoted by $\lambda(k,t)$, is given by
$
\lambda(k,t) =
\rm{kmax}_{(b,d)\in D}
\{\rm{max}\{0, min\{t-b,d-t\}\}
\}
$ 
where $\rm{kmax}$ denotes the $k$th largest value, see Figure~\ref{fig:PH}(b) ii.

This approach is applied to two point clouds extracted from the CT images: the arterial point cloud, consisting of all vertices along the vascular centerline, and the complement point cloud. These constructions provide complementary representations of vascular geometry and the spatial organization of the surrounding regions. The analysis is performed on both the full vascular network and each lobe individually.

\begin{figure}
\centering
\includegraphics[width=\linewidth]{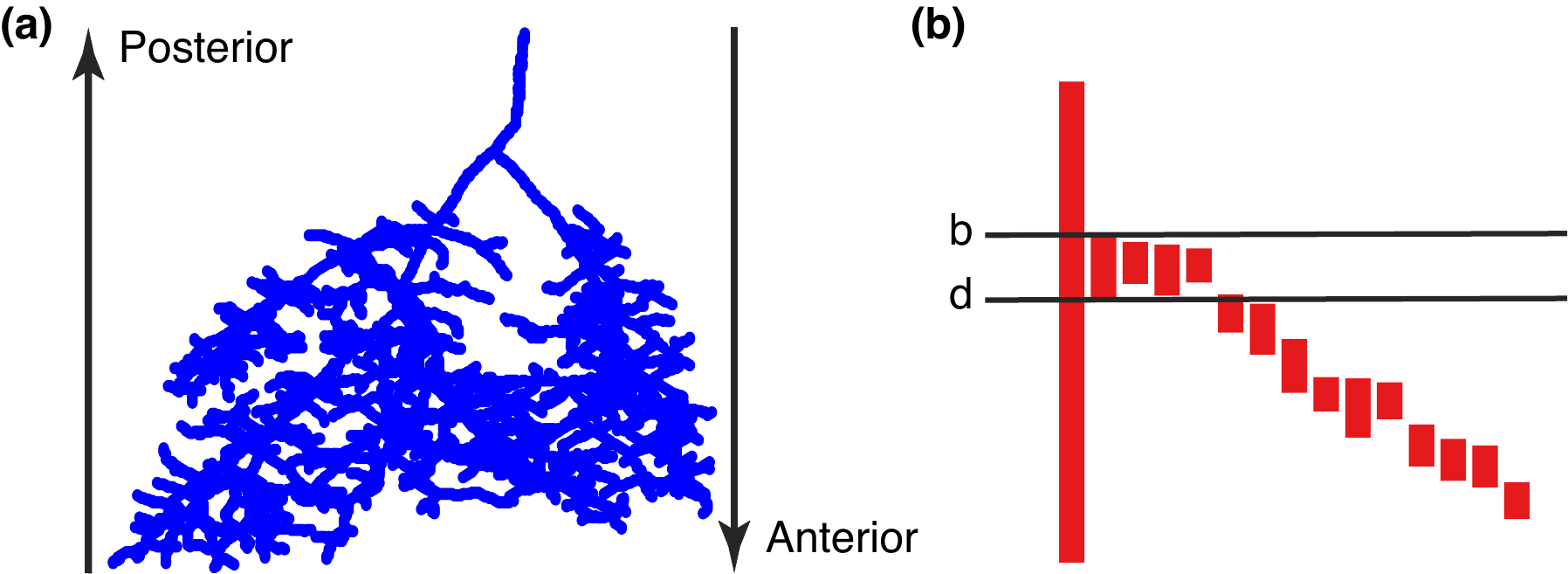}
\caption{(a) Mouse network with labeled directions for height filtration. (b) Persistence barcode with the 15 longest-living generators from an anterior filtration. \textit{t=b} is the generator's birth time, and 
\(t=d\) is the death time at which the component merges with an older component; a component that never merges is essential and has infinite death time.}
\label{fig:tdaMethods}
\end{figure}

\medskip
\noindent\textbf{Directional Complexity.}  Following the height-filtration framework of Belchi et al.~\cite{Brodzki18} and Miller et al.~\cite{Miller23}, we quantify directional changes in the arterial tree using a Morse-theory-inspired height filtration. Let $X\subset\mathbb{R}^3$ denote the arterial tree represented as a one-dimensional simplicial complex, and let $v\in S^2$ denote a filtration direction. The height function
$h_v(x)=\langle x,v\rangle$
induces the filtration 
$X_t=\{\sigma\in X:\max_{x\in\sigma}h_v(x)\leq t\}.$ 
We compute $0$-dimensional persistent homology for filtrations in the positive and negative $z$-directions, corresponding to posterior and anterior sweeps, respectively; see Figure~\ref{fig:tdaMethods}(a).

For each direction, let $D_0^v$ denote the resulting $0$- dimensional persistence diagram. We record both the number of finite persistence intervals and the \textit{directional complexity}, defined as the total persistence
$\mathrm{DC}(v) =
\sum_{\substack{(b,d)\in D_0^v\\ d<\infty}}
(d-b).$
The number of finite intervals records how many $H_0$ features are created and subsequently merged during the directional sweep, whereas directional complexity additionally incorporates their persistence. Thus, two arterial trees may produce a similar number of $H_0$ features while differing in directional complexity if those features persist over different ranges of the filtration.

Equivalently, directional complexity is the area under the corresponding $0$-dimensional Betti curve,
$\mathrm{DC}(v) =
\int \beta_0(D_0^v,t)\,dt.$
This relationship motivates our additional use of persistence curves and landscapes, which retain information about how the topological features are distributed across the filtration domain rather than reducing the persistence diagram to a single scalar summary. In particular, the Betti curve records the number of active $H_0$ features at each filtration value, while the lifespan curve weights each active feature by its persistence. Together, these summaries provide complementary descriptions of the directional branching structure of the arterial tree~\cite{Brodzki18,Miller23}; see Figure~\ref{fig:tdaMethods}(b).
\medskip

\noindent\textbf{Statistical Comparison of Persistence Curves.} Persistence curves are treated as functional data and compared between groups. For each curve type, homological dimension, and anatomical region, curves are evaluated on a common filtration grid to ensure alignment across subjects. Following Aslam et al.~\cite{Aslam2022}, group differences are assessed using a permutation test with the $L_2$ norm of the difference between the group mean curves as the test statistic. These tests are performed by permuting the control and hypertensive group labels across mice while preserving the original group sizes, and recomputing the $L_2$ distance between the resulting group mean curves for each permutation. The permutation p-value is then calculated as the proportion of permuted $L_2$ distances that are at least as large as the observed $L_2$ distance. This provides a global test of whether the persistence curves differ between groups over the filtration domain. Tests are performed separately for each curve type, homological dimension, and anatomical region.

\subsection{Statistical Shape Analysis}\label{Msec:SSA}
\smallskip

\begin{figure*}
    \centering
    \includegraphics[width=\linewidth]{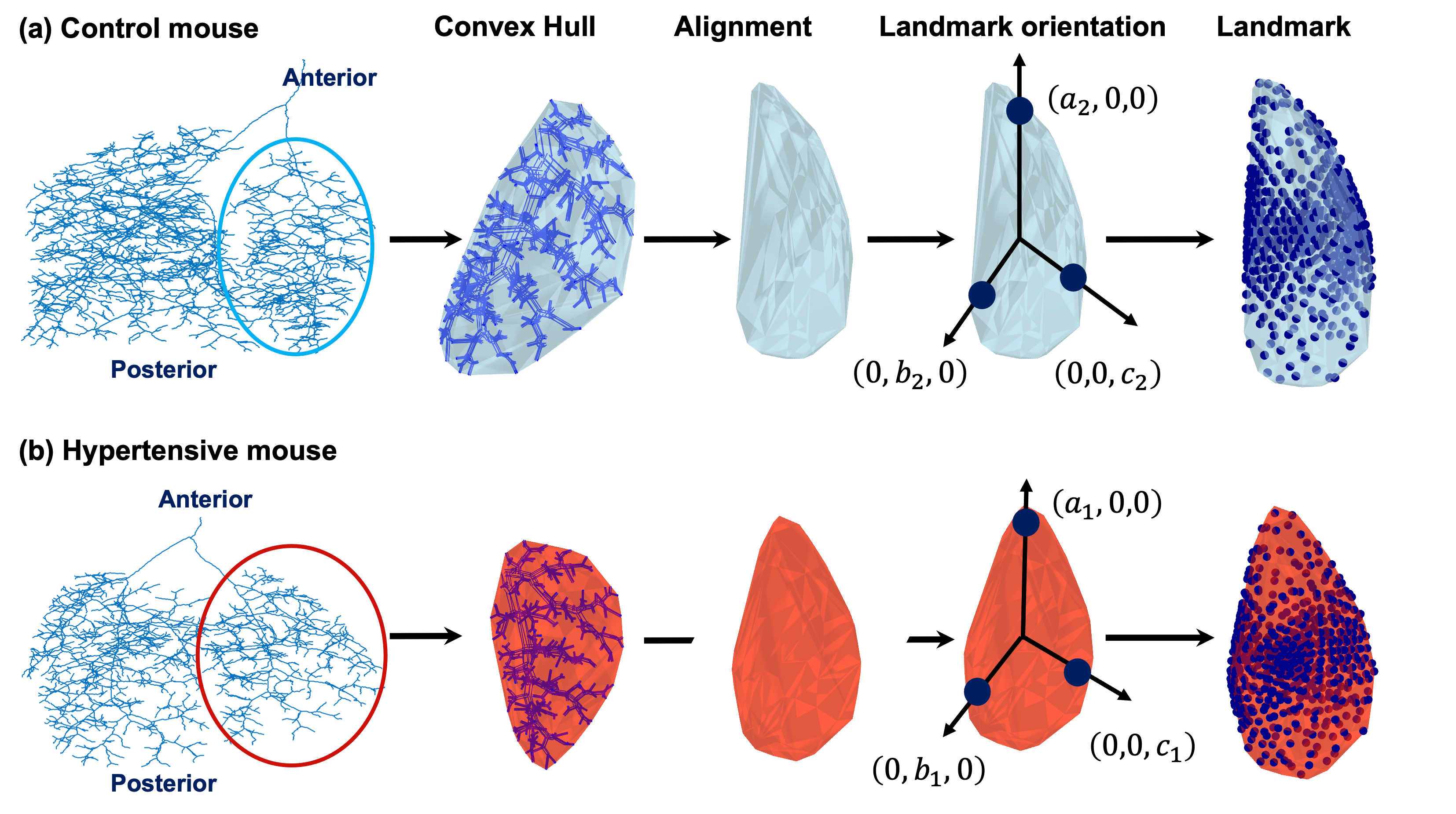}
    \caption{The pipeline for generating landmarks is illustrated for the left lung in a representative control (a) and hypertensive (b) mouse. First, centerline points for each lobe are extracted, and convex hulls are constructed using the triangulation algorithm. Next, we align the lobes regardless of treatment group. Lastly, arbitrarily chosen vectors (pictured: elementary vectors) are used to identify points on the convex hull surface, which serve as landmarks. In practice, we use around 2000 landmarks (see Supplementary Materials Section 1.2.1).}
    \label{fig:ssa}
\end{figure*}

\noindent Statistical shape analysis (SSA) provides an approach to comparing the \textit{shape} of lung lobes of control and hypertensive mice \cite{drydenmardiaStatisticalShapeAnalysis2016}. We perform SSA on the convex hulls of each mouse lung lobe, comparing the outside shapes of the lung regardless of their branching patterns. SSA uses a collection of landmarks (points of correspondence on each object) to describe an object's shape after standardizing for orientation, shift, and size. Figure~\ref{fig:ssa} shows the pipeline for calculating the landmarks for each mouse lobe. We employ standard SSA procedures based on principal component analysis (PCA) on a collection of $k$ landmarks across our sample to determine whether shape differs significantly between control and hypertensive mice. 

\medskip
\noindent{\bf Alignment.} Before defining landmarks, we align convex hulls spatially (see Figure~\ref{fig:ssa}) using a two-step procedure consisting of an initial coarse alignment generated using the Iterative Closest Point (ICP) algorithm \cite{Besl92}, and a refinement step obtained using the Coherent Point Drift (CPD) algorithm \cite{Myronenko10}. 

First, we select a reference lobe network and extract its points to form a 3D point cloud. PCA is then performed on this reference cloud to extract the first principal component (PC1), which captures the network's dominant geometric direction. Using the Rodrigues rotation formula \cite{Rodrigues1840, Valdenebro2016}, PC1 is rigidly aligned with the $z$-axis, establishing a consistent reference orientation. Next, a set of point clouds corresponding to the same lobe from a different mouse is selected for alignment to the reference cloud. In the first alignment stage, MATLAB’s Computer Vision Toolbox function \texttt{pcregistericp} \cite{MATLAB:pcregistericp, Besl92} is used to sequentially register each unaligned point cloud to the reference using the rigid transformation option. This step produces a coarse alignment minimizing the point-to-point distances between the point clouds. 

In the second stage, the alignment is refined using the CPD algorithm implemented in MATLAB’s \texttt{pcregistercpd} \cite{MATLAB:pcregistercpd, Myronenko10} function. CPD treats the moving point cloud as a Gaussian mixture model and estimates the rigid transformation that best aligns it with the reference cloud, preserving its shape. The final alignment error is quantified from bidirectional nearest-neighbor distances between the two clouds, summarized as the mean, median, RMS (Chamfer), and maximum (Hausdorff) distances.

\medskip
\noindent\textbf{Sampling Algorithms.} Convex hulls do not have visually identifiable landmarks. To use SSA, we define landmarks that correspond across convex hulls of different lobe networks. These landmarks denote the intersections between a vector and the surface of the convex hull centered at the origin (see Figure~\ref{fig:ssa}). This is done using a predefined collection of $k$ vectors that make up the $k$ corresponding landmarks for each convex hull. In the analysis, we set $k$ to an arbitrarily large number, only limited by computational burden. In the Supplementary Materials Section 1.2.1, we show that the results are robust to the choice of sampling algorithm. 

\medskip
\noindent\textbf{Classification with SSA Features.} We perform tangent-based shape PCA \cite{drydenmardiaStatisticalShapeAnalysis2016} to compare the shape of the convex hulls between the hypertensive and control mice. Shape PCA vectorizes the coordinates of the $k$ landmarks and projects them onto the tangent space to the Riemannian manifold of shapes. Then, PCA is applied after standardization of the tangent space. Details of the algorithm are provided in the Supplementary Materials Section 1.2.2. Once we obtain the principal component scores for the dataset, we create a scatterplot of the first and second PC scores to visualize the relationship between control and treatment convex hull shapes across different pressures. We use support vector machines (SVMs) to assess the ability to diagnose pulmonary hypertension using convex hull shapes. 

\subsection{Classification of Hypertensive Pulmonary Arterial Networks}
\smallskip

\noindent To assess whether the extracted features can identify pulmonary hypertension in mice, we construct several classification models. We combine the geometric summaries computed in Section \ref{Msec:GeomFeat}, persistent homology summaries via persistence curves from Section \ref{Msec:PH}, and the shape principal component scores from Section \ref{Msec:SSA} in a classification model. The persistent homology summaries we use are detailed in the Supplementary Materials Section 1.4. 
 
We model the categorical hypertension variable at three levels: (1) the complete lung, computing features over the entire pulmonary arterial network; (2) the ``Tall-Lobe" level, computing features individually from each lobe, using lobe type as an additional categorical feature; and (3) a ``Wide-Lobe" level, computing features separately on the lobe level, treating them as independent variables. At each level, we consider three different classification models of the features. The first is a linear Bayesian spike-and-slab model to assess whether each feature is present when predicting pulmonary hypertension. Details on the MCMC implementation of the Bayesian model are in the Supplementary Materials Section 1.3. 

The second is a boosted tree model from the R package \texttt{gpboost} \cite{sigrist2021gpboost}. We compare the linear parametrization to the boosted tree to see if the features have a nonlinear relationship with pulmonary hypertension. This package allows us to add random effects to account for dependence within each mouse and provide feature importance for each predictor, which we will compare with the posterior inclusion probabilities from the Bayesian Sparse Linear Regression results. Our third considered classification model is the support vector machine (SVM) from the R package \texttt{e1071} \cite{e1071_R_Package}.

\section{Results}

\subsection{Geometric Features}\label{sec:geom}
\smallskip 

\begin{figure*}[h!tbp]
\centering
\includegraphics[width=0.8\textwidth]{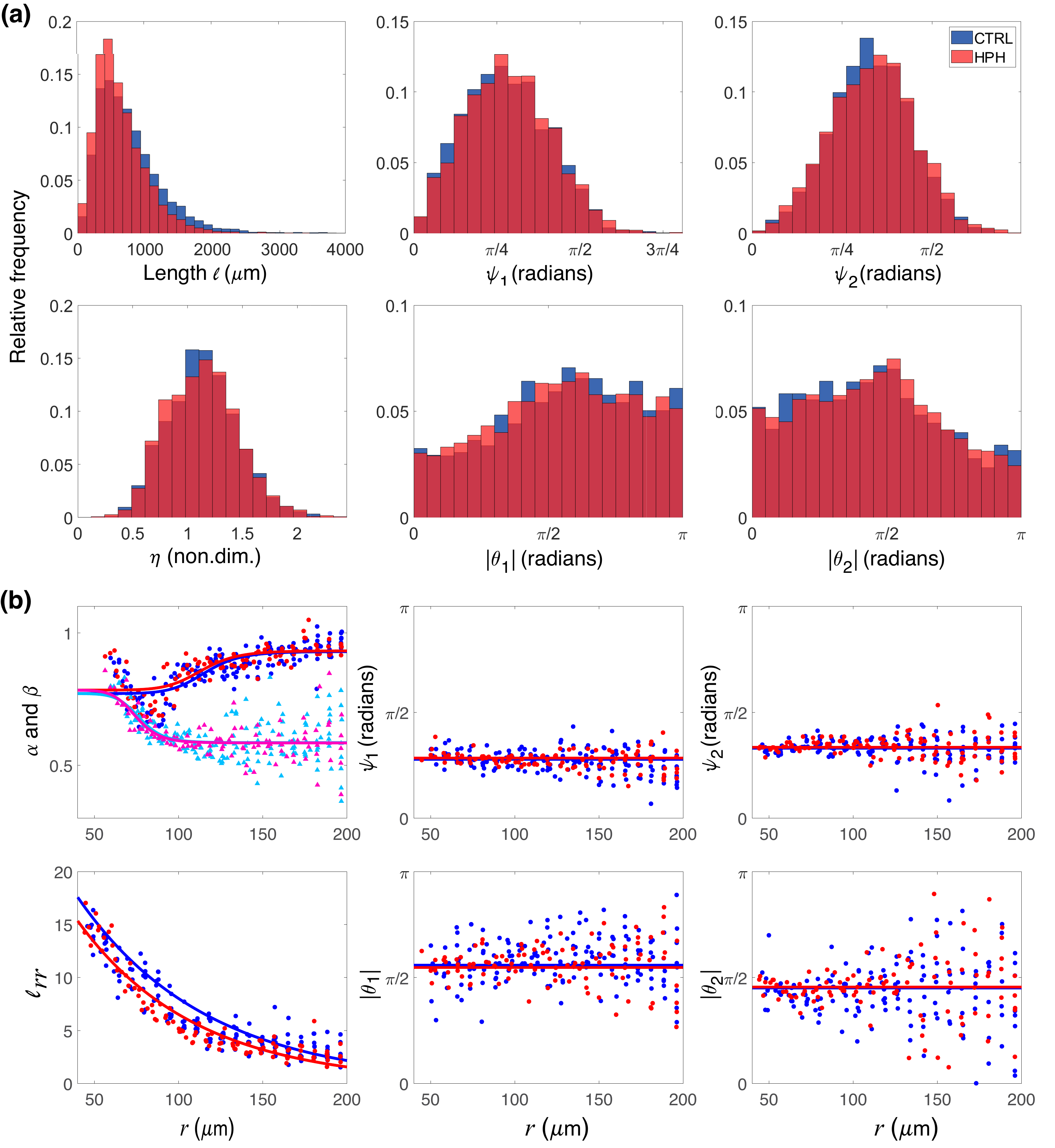}
       \caption{Geometric features, functions, and distributions at 6.3 mmHg, with HPH in red/green and CTRL in blue/purple. (a) Histograms of length, $\eta$, and the four angles. (b) Curve fits to $\alpha$ and $\beta$ (sigmoid), an exponential fit to $\ell_{rr}$, and average values of the angles.}
    \label{fig:geomFigures}
\end{figure*}
\noindent Across all lobes, the median radius $r$ and length-to-radius ratio $\ell_{rr}$ significantly differ between the control and hypertensive animals. Results listed in Table \ref{tab:average_features} comparing average values of the geometric features in the pruned trees show that $r$ is significantly larger in HPH animals. At the same time,  $\ell_{rr}$ is smaller in HPH animals. Examination of Figure~\ref{fig:geomFigures}(b) shows that differences in $\ell_{rr}$ are larger in small vessels, agreeing with the notion that remodeling starts in the small arteries and then migrates to the large vessels. This finding aligns with the vessel lengths $\ell$ being shorter in HPH, with the difference being significant in the full network and the middle and post-caval lobes. Further, Figure~\ref{fig:geomFigures}(a) depicts a histogram of vessel lengths $\ell$, and we see that the HPH group has a higher average count across all mice at lower vessel lengths.

In fact, we find that the average values of the radius scaling parameters $\alpha$, $\gamma$,  $\eta$, and $\xi$ also differ significantly between the two groups in all lobes. Table \ref{tab:average_features} provides the average geometric feature values in both the hypertensive and control groups at 6.3 mmHg, since this pressure maximizes the expected differences between groups. This table shows that $\alpha$ is higher in the hypertensive animals. A higher $\alpha$ generates structured trees with more vessels, providing higher flow resistance, again consistent with the higher sustained pressure in the hypertensive group. Yet comparison of $\alpha$ and $\beta$ (Figure~\ref{fig:geomFigures}(b)) show that the functions of radius do not differ significantly between the two groups. As stated by equation (\ref{eq:alphabetaGamma}), $\gamma$ is inversely proportional to $\alpha^2$, thus, the observed decrease in $\gamma$ in HPH animals is consistent with the corresponding increase in $\alpha$. Moreover, since $\eta = \alpha^2 + \beta^2$, an increase in $\alpha$ necessarily contributes to an increase in $\eta$, consistent with the elevated $\eta$ observed in HPH mice, also shown in Figure~\ref{fig:geomFigures}(a). Table \ref{tab:average_features} also shows that $\xi$ increases significantly in all lobes and in the full lung. As expected, these results agree with findings for $\alpha$, $\gamma$, $\eta$, and $\xi$, though it is worth noting (see equation \ref{eq:alphabetaGamma}) that only three of these parameters are independent \cite{Olufsen98}. Averages for the other three pressures, as well as $p$-values from t-tests performed on the geometric features, can be found in Tables S2-S5 in the Supplementary Materials.

Tortuosity $(\kappa)$ decreases in HPH mice, with the difference being significant across the full network. Conversely, $\psi_2$ is higher in HPH mice, with the difference being significant in the inferior lobe and the full network.  The distribution of angles can be directly observed in Figure~\ref{fig:geomFigures}(a). The in-plane angles ($\psi_i$) are concentrated primarily between $\pi/4$ and $\pi/2$, with the out-of-plane angles reported as absolute values ($|\theta_i|$) concentrated primarily between $\pi/4$ and $3\pi/4$. The original $\theta_i$ values followed a bimodal distribution on $[-\pi,\pi]$ (see Supplementary Materials Figure~S8).

The ANCOVA results, with the related t-values and p-values for both the median and the standard deviation, are in Supplementary Materials Tables S6-S9. The standard deviation of $\alpha$ decreased consistently in HPH mice, while it increased for $\beta$. 

\begin{table*}[htbp]
\centering
\footnotesize
\caption{Average geometric features for blood vessels in each mouse lung lobe at 6.3 mmHg. Raw vessel-segment measurements were first averaged within each mouse, then group means $\pm$ SD were computed across mice. Black values correspond to the control group ($n=8$ mice), and \textcolor{red}{red} values correspond to the hypertensive group ($n=5$ mice). Values in \textbf{bold} differ significantly ($p<0.05$) in the Welch's t-test. P-values can be found in the Supplementary Materials Tables S3 and S5.}
\label{tab:average_features}
\begin{tabular}{c|ccccc}
\toprule
\textbf{Feature} & \textbf{Left} & \textbf{Superior} & \textbf{Middle} & \textbf{Inferior} & \textbf{Post-caval} \\
\midrule

\multirow{2}{*}{$r$ ($\mu$m)}
& $\mathbf{97.3\pm6.9}$ & $\mathbf{99.4\pm6.1}$ & $\mathbf{92.8\pm8.0}$ & $\mathbf{99.1\pm7.6}$ & $\mathbf{95.4\pm5.1}$ \\
& \textcolor{red}{$\mathbf{112\pm3.8}$} & \textcolor{red}{$\mathbf{110\pm3.6}$} & \textcolor{red}{$\mathbf{106\pm2.9}$} & \textcolor{red}{$\mathbf{112\pm4.0}$} & \textcolor{red}{$\mathbf{106\pm4.0}$} \\

\multirow{2}{*}{$\ell$ ($\mu$m)}
& $\mathbf{764\pm68}$ & $\mathbf{760\pm110}$ & $\mathbf{815\pm55}$ & $\mathbf{762\pm91}$ & $\mathbf{870\pm110}$ \\
& \textcolor{red}{$\mathbf{670\pm61}$} & \textcolor{red}{$\mathbf{633\pm49}$} & \textcolor{red}{$\mathbf{666\pm47}$} & \textcolor{red}{$\mathbf{667\pm40}$} & \textcolor{red}{$\mathbf{676\pm68}$} \\

\multirow{2}{*}{$V$ ($10^{7}\,\mu$m$^3$)}
& $3.40\pm0.74$ & $2.95\pm0.78$ & $2.84\pm0.53$ & $3.17\pm0.74$ & $3.21\pm0.66$ \\
& \textcolor{red}{$3.59\pm0.28$} & \textcolor{red}{$2.89\pm0.16$} & \textcolor{red}{$2.85\pm0.16$} & \textcolor{red}{$3.30\pm0.28$} & \textcolor{red}{$3.01\pm0.47$} \\

\multirow{2}{*}{$\alpha$}
& $\mathbf{0.845\pm0.022}$ & $\mathbf{0.833\pm0.025}$ & $\mathbf{0.840\pm0.020}$ & $\mathbf{0.834\pm0.008}$ & $\mathbf{0.840\pm0.023}$ \\
& \textcolor{red}{$\mathbf{0.925\pm0.019}$} & \textcolor{red}{$\mathbf{0.927\pm0.017}$} & \textcolor{red}{$\mathbf{0.934\pm0.017}$} & \textcolor{red}{$\mathbf{0.933\pm0.019}$} & \textcolor{red}{$\mathbf{0.939\pm0.015}$} \\

\multirow{2}{*}{$\beta$}
& $0.620\pm0.024$ & $0.591\pm0.026$ & $0.596\pm0.021$ & $0.607\pm0.019$ & $0.595\pm0.028$ \\
& \textcolor{red}{$0.629\pm0.014$} & \textcolor{red}{$0.590\pm0.030$} & \textcolor{red}{$0.590\pm0.029$} & \textcolor{red}{$0.614\pm0.026$} & \textcolor{red}{$0.590\pm0.017$} \\

\multirow{2}{*}{$\xi$}
& $\mathbf{2.44\pm0.19}$ & $\mathbf{2.25\pm0.26}$ & $\mathbf{2.25\pm0.15}$ & $\mathbf{2.24\pm0.14}$ & $\mathbf{2.25\pm0.32}$ \\
& \textcolor{red}{$\mathbf{2.90\pm0.27}$} & \textcolor{red}{$\mathbf{2.69\pm0.24}$} & \textcolor{red}{$\mathbf{2.68\pm0.19}$} & \textcolor{red}{$\mathbf{2.98\pm0.27}$} & \textcolor{red}{$\mathbf{2.70\pm0.05}$} \\

\multirow{2}{*}{$\gamma$}
& $\mathbf{0.594\pm0.030}$ & $\mathbf{0.563\pm0.059}$ & $\mathbf{0.563\pm0.025}$ & $\mathbf{0.589\pm0.033}$ & $\mathbf{0.565\pm0.044}$ \\
& \textcolor{red}{$\mathbf{0.501\pm0.023}$} & \textcolor{red}{$\mathbf{0.452\pm0.042}$} & \textcolor{red}{$\mathbf{0.439\pm0.039}$} & \textcolor{red}{$\mathbf{0.480\pm0.046}$} & \textcolor{red}{$\mathbf{0.429\pm0.032}$} \\

\multirow{2}{*}{$\eta$}
& $\mathbf{1.14\pm0.06}$ & $\mathbf{1.09\pm0.04}$ & $\mathbf{1.10\pm0.05}$ & $\mathbf{1.11\pm0.03}$ & $\mathbf{1.11\pm0.05}$ \\
& \textcolor{red}{$\mathbf{1.28\pm0.05}$} & \textcolor{red}{$\mathbf{1.25\pm0.07}$} & \textcolor{red}{$\mathbf{1.26\pm0.04}$} & \textcolor{red}{$\mathbf{1.28\pm0.05}$} & \textcolor{red}{$\mathbf{1.27\pm0.03}$} \\

\multirow{2}{*}{$\psi_1$ (rad)}
& $0.882\pm0.034$ & $0.870\pm0.066$ & $0.846\pm0.064$ & $0.865\pm0.050$ & $0.855\pm0.077$ \\
& \textcolor{red}{$0.892\pm0.024$} & \textcolor{red}{$0.839\pm0.044$} & \textcolor{red}{$0.870\pm0.049$} & \textcolor{red}{$0.867\pm0.059$} & \textcolor{red}{$0.879\pm0.090$} \\

\multirow{2}{*}{$\psi_2$ (rad)}
& $\mathbf{1.04\pm0.02}$ & $\mathbf{1.03\pm0.04}$ & $1.08\pm0.07$ & $\mathbf{1.03\pm0.04}$ & $1.06\pm0.06$ \\
& \textcolor{red}{$\mathbf{1.08\pm0.02}$} & \textcolor{red}{$\mathbf{1.12\pm0.06}$} & \textcolor{red}{$1.09\pm0.03$} & \textcolor{red}{$\mathbf{1.10\pm0.04}$} & \textcolor{red}{$1.10\pm0.08$} \\

\multirow{2}{*}{$\theta_1$ (rad)}
& $0.01\pm0.19$ & $0.00\pm0.20$ & $0.03\pm0.49$ & $0.02\pm0.25$ & $0.15\pm0.32$ \\
& \textcolor{red}{$0.06\pm0.06$} & \textcolor{red}{$0.02\pm0.24$} & \textcolor{red}{$-0.07\pm0.17$} & \textcolor{red}{$0.09\pm0.16$} & \textcolor{red}{$-0.16\pm0.34$} \\

\multirow{2}{*}{$\theta_2$ (rad)}
& $0.05\pm0.16$ & $-0.02\pm0.26$ & $-0.02\pm0.25$ & $-0.04\pm0.21$ & $\mathbf{-0.17\pm0.27}$ \\
& \textcolor{red}{$0.05\pm0.12$} & \textcolor{red}{$0.02\pm0.11$} & \textcolor{red}{$-0.05\pm0.20$} & \textcolor{red}{$-0.05\pm0.10$} & \textcolor{red}{$\mathbf{0.11\pm0.18}$} \\

$D$ & \shortstack{$10.6\pm1.6$ \\ \textcolor{red}{$11.1\pm0.9$}} & \shortstack{$7.90\pm1.60$ \\ \textcolor{red}{$8.58\pm0.54$}} & \shortstack{$7.14\pm0.68$ \\ \textcolor{red}{$7.97\pm0.93$}} & \shortstack{$10.8\pm1.0$ \\ \textcolor{red}{$11.8\pm1.1$}} & \shortstack{$7.5\pm1.0$ \\ \textcolor{red}{$8.33\pm0.88$}} \\

\multirow{2}{*}{$L_R$ ($\mu$m)}
& $8900\pm660$ & $5380\pm540$ & $5110\pm630$ & $8700\pm1300$ & $5470\pm570$ \\
& \textcolor{red}{$8230\pm830$} & \textcolor{red}{$5320\pm230$} & \textcolor{red}{$4820\pm550$} & \textcolor{red}{$8190\pm960$} & \textcolor{red}{$5430\pm540$} \\

\multirow{2}{*}{$\kappa$}
& $\mathbf{1.21\pm0.01}$ & $1.21\pm0.01$ & $\mathbf{1.21\pm0.02}$ & $1.21\pm0.02$ & $1.23\pm0.03$ \\
& \textcolor{red}{$\mathbf{1.20\pm0.07}$} & \textcolor{red}{$1.19\pm0.02$} & \textcolor{red}{$\mathbf{1.19\pm0.01}$} & \textcolor{red}{$1.21\pm0.01$} & \textcolor{red}{$1.25\pm0.04$} \\

\multirow{2}{*}{$\ell_{rr}$}
& $\mathbf{9.43\pm0.85}$ & $\mathbf{9.46\pm0.75}$ & $\mathbf{10.6\pm0.9}$ & $\mathbf{9.59\pm0.75}$ & $\mathbf{10.8\pm1.2}$ \\
& \textcolor{red}{$\mathbf{7.35\pm 0.81}$} &   
  \textcolor{red}{$\mathbf{7.38\pm 0.80}$} & \textcolor{red}{$\mathbf{7.83\pm 0.60}$} & \textcolor{red}{$\mathbf{7.51\pm 0.65}$} & \textcolor{red}{$\mathbf{7.70\pm 0.58}$} \\
\bottomrule
\end{tabular}
\end{table*}

\begin{equation*}
    \ell_{rr} = C_1 e^{-C_2 r}.
\end{equation*}
$\alpha$ and $\beta$, which can be fitted to sigmoidal curves on the form 
\begin{align*}
    S_\alpha(r) &= (\alpha_M - \alpha_m)\frac{r^k}{r_{0\alpha}^k + r^k} + \alpha_m \\ 
    S_\beta(r) &= (\beta_M-\beta_m)\frac{r_{0\beta}^k}{r_{0\beta}^k+r^k}+\beta_m,
\end{align*}
where $\alpha_m = \beta_M$ and $k$ is the same for both equations (Figure~\ref{fig:geomFigures}(b)). Fitting the data yields the coefficients in Table \ref{tab:geomcoef}. $C_1$ decreases in HPH mice. $\alpha_M$ and $k$ increase overall at higher pressures, while $\beta_m$ decreases overall. The rest of the coefficients did not differ noticeably between groups. 

\begin{table*}[ht]
\centering
\caption{Coefficient information, control values in black and hypertensive in \textcolor{red}{red}.}
\begin{tabular}{c c c c c }
\toprule
\textbf{} & \textbf{6.3 mmHg} & \textbf{7.4 mmHg} & \textbf{13.0 mmHg} & \textbf{17.2 mmHg} \\
\midrule
\multirow{2}{*}{$C_1$}
& $22.2$ & $20.2$ & $18.2$ & $17.2$ \\
& \textcolor{red}{$18.5$} & \textcolor{red}{$17.3$} & \textcolor{red}{$17.3$} & \textcolor{red}{$14.5$} \\[0.15cm]
\multirow{2}{*}{$C_2$}
& $0.0112$ & $0.0106$ & $0.0108$ & $0.0110$ \\
& \textcolor{red}{$0.0112$} & \textcolor{red}{$0.0108$} & \textcolor{red}{$0.0110$} & \textcolor{red}{$0.0099$} \\[0.15cm]
\multirow{2}{*}{$\alpha_M$}
& $ 0.929 $ & $ 0.932 $ & $ 0.911 $ & $ 0.897 $ \\
& \textcolor{red}{$ 0.932 $} & \textcolor{red}{$ 0.916 $} & \textcolor{red}{$ 0.921 $} & \textcolor{red}{$ 0.899 $} \\[0.15cm]
\multirow{2}{*}{$\beta_m$}
& $ 0.585 $ & $ 0.584 $ & $ 0.597 $ & $ 0.602 $ \\
& \textcolor{red}{$ 0.584 $} & \textcolor{red}{$ 0.601 $} & \textcolor{red}{$ 0.595 $} & \textcolor{red}{$ 0.612 $} \\[0.15cm]
\multirow{2}{*}{$\alpha_m=\beta_M$}
& $0.771$ & $0.792$ & $0.786$ & $0.773$ \\
& \textcolor{red}{$0.784$} & \textcolor{red}{$0.778$} & \textcolor{red}{$0.796$} & \textcolor{red}{$0.805$} \\[0.15cm]
\multirow{2}{*}{$r_{0\alpha}$}
& $116$ & $132$& $118$ & $115$ \\
& \textcolor{red}{$114$} & \textcolor{red}{$106$} & \textcolor{red}{$118$} & \textcolor{red}{$104$} \\[0.15cm]
\multirow{2}{*}{$r_{0\beta}$}
& $76.8$ & $78.6$ & $78.6$ & $76.0$ \\
& \textcolor{red}{$74.5$} & \textcolor{red}{$75.3$} & \textcolor{red}{$76.4$} & \textcolor{red}{$76.7$} \\[0.15cm]
\multirow{2}{*}{$k$}
& $11.0$ & $10.1$ & $12.9$ & $16.7$ \\
& \textcolor{red}{$10.6$} & \textcolor{red}{$14.6$} & \textcolor{red}{$13.0$} & \textcolor{red}{$14.0$} \\
\bottomrule
\end{tabular}
\label{tab:geomcoef}
\end{table*}

\subsection{Persistent Homology}
\smallskip

\noindent\textbf{Directional Complexity.}
The height-filtration summaries are reported in Table~\ref{tab:dcResults}. Mean directional complexity was lower in HPH mice than in CTRL mice in both the anterior and posterior directions at all four pressures. The largest whole-network separation occurred in the posterior direction at 6.3 mmHg, where mean directional complexity was $197{,}853 \pm 30{,}207$ in CTRL mice and $160{,}112 \pm 24{,}548$ in HPH mice, corresponding to a reduction of approximately $19\%$ in HPH. This difference was statistically significant ($p=0.033$). Although directional complexity generally increased with pressure in both groups, the magnitude of the difference between CTRL and HPH did not vary monotonically with pressure.
\begin{table*}[ht]
\centering
\caption{Average and standard deviations for the features extracted from the height filtration. Control values are shown in black and hypertensive values in \textcolor{red}{red}. Directional Complexity at 6.3 mmHg in the posterior direction was the only significant difference ($p = 0.033$).}
\label{tab:dcResults}
\begin{tabular}{ccccc}
\toprule
& \multicolumn{2}{c}{\textbf{Anterior}} &
\multicolumn{2}{c}{\textbf{Posterior}} \\
\cmidrule(lr){2-3} \cmidrule(lr){4-5}
\textbf{Pressure} &
\makecell{\textbf{Directional}\\\textbf{Complexity}} & \textbf{$H_0$ feature count} &
\makecell{\textbf{Directional}\\\textbf{Complexity}} & \textbf{$H_0$ feature count} \\
\midrule
6.3 mmHg &
$71857 \pm 15433$ & $178 \pm 41$ &
$\mathbf{197853 \pm 30207}$ & $245 \pm 42$ \\
&
\textcolor{red}{$68623 \pm 8192$} &
\textcolor{red}{$191 \pm 5$} &
\textcolor{red}{$\mathbf{160112 \pm 24548}$} &
\textcolor{red}{$227 \pm 14$} \\
\addlinespace

7.4 mmHg &
$75683 \pm 16426$ & $194 \pm 46$ &
$175488 \pm 67963$ & $253 \pm 53$ \\
&
\textcolor{red}{$71745 \pm 8243$} &
\textcolor{red}{$205 \pm 8$} &
\textcolor{red}{$163448 \pm 25667$} &
\textcolor{red}{$239 \pm 11$} \\
\addlinespace

13.0 mmHg &
$113977 \pm 24994$ & $301 \pm 57$ &
$255994 \pm 32724$ & $371 \pm 59$ \\
&
\textcolor{red}{$112216 \pm 12102$} &
\textcolor{red}{$307 \pm 7$} &
\textcolor{red}{$236430 \pm 25185$} &
\textcolor{red}{$364 \pm 9$} \\
\addlinespace

17.2 mmHg &
$163851 \pm 24605$ & $444 \pm 47$ &
$322884 \pm 36152$ & $531 \pm 47$ \\
&
\textcolor{red}{$156640 \pm 12895$} &
\textcolor{red}{$444 \pm 33$} &
\textcolor{red}{$288078 \pm 33973$} &
\textcolor{red}{$506 \pm 38$} \\
\bottomrule
\end{tabular}
\end{table*}

The number of finite $H_0$ persistence intervals showed a different pattern between the two filtration directions. In the anterior direction, HPH mice had a higher mean number of finite intervals than CTRL mice at 6.3, 7.4, and 13.0 mmHg, while the two groups had the same mean count at 17.2 mmHg. In contrast, in the posterior direction, HPH mice had fewer finite intervals than CTRL mice at all four pressures. The anterior differences in interval count were not statistically significant. Thus, the differences in directional complexity were not consistently accompanied by corresponding differences in the number of finite $H_0$ intervals.

Since directional complexity is the total persistence of the finite $H_0$ intervals, equivalently the area under the corresponding Betti curve, we further examine the functional persistence summaries to determine how this persistence is distributed across the filtration domain. The Betti, lifespan, normalized lifespan, and persistence-landscape curves revealed additional pressure- and lobe-specific differences.

In the inferior lobe, posterior Betti and lifespan curves differed significantly between CTRL and HPH mice at both 6.3 and 17.2 mmHg, with CTRL mice exhibiting larger Betti and lifespan values. In the post-caval lobe, the anterior $\lambda_3$ persistence landscape differed significantly between groups at all four pressures ($p=0.032$, $0.041$, $0.013$, and $0.006$ at 17.2, 13.0, 7.4, and 6.3 mmHg, respectively), with longer decay time in HPH mice. An additional significant difference in the anterior Betti curve was detected in the post-caval lobe at 6.3 mmHg.

In contrast, no significant group differences were detected in the full-network persistence curves or persistence landscapes after pruning, and the normalized lifespan curves showed little evidence of group separation. Full permutation-test results for the height-filtration persistence summaries are provided in Supplementary Materials Table~S15. Overall, the height-filtration analysis identified a significant whole-network difference in posterior directional complexity at 6.3 mmHg, together with additional pressure- and lobe-specific differences in the functional persistence summaries.

\begin{table*}[ht]
\centering
\setlength{\tabcolsep}{4pt}
\caption{Branch-to-branch proximity values (mean $\pm$ SD) for CTRL and HPH mice. P denotes pressure (mmHg). Values in \textbf{bold} differ significantly ($p<0.05$) based on a t-test. P-values can be found in Supplementary Materials Table S10.}
\begin{tabular}{c c c c c c c}
\toprule
\textbf{\shortstack{P}} &
\textbf{Full} &
\textbf{Left} &
\textbf{Superior} &
\textbf{Middle} &
\textbf{Inferior} &
\textbf{Post-caval} \\
\midrule

\multirow{2}{*}{6.3}
& $\mathbf{5.21\pm0.47}$
& $\mathbf{2.17\pm0.22}$
& $1.03\pm0.22$
& $1.16\pm0.30$
& $1.72\pm0.37$
& $\mathbf{0.783\pm0.13}$ \\
& \textcolor{red}{$\mathbf{4.71\pm0.33}$}
& \textcolor{red}{$\mathbf{1.78\pm0.29}$}
& \textcolor{red}{$0.911\pm0.161$}
& \textcolor{red}{$0.964\pm0.130$}
& \textcolor{red}{$1.45\pm0.06$}
& \textcolor{red}{$\mathbf{0.650\pm0.065}$} \\[0.15cm]

\multirow{2}{*}{7.4}
& $\mathbf{5.52\pm0.45}$
& $2.33\pm0.28$
& $1.07\pm0.18$
& $1.19\pm0.25$
& $1.79\pm0.29$
& $\mathbf{0.816\pm0.11}$ \\

& \textcolor{red}{$\mathbf{4.91\pm0.26}$}
& \textcolor{red}{$1.99\pm0.28$}
& \textcolor{red}{$0.998\pm0.153$}
& \textcolor{red}{$1.01\pm0.06$}
& \textcolor{red}{$1.46\pm0.24$}
& \textcolor{red}{$\mathbf{0.614\pm0.126}$} \\[0.15cm]

\multirow{2}{*}{13.0}

& $6.33\pm0.89$
& $2.57\pm0.35$
& $1.36\pm0.25$
& $1.61\pm0.59$
& $1.97\pm0.56$
& $0.911\pm0.174$ \\

& \textcolor{red}{$6.07\pm0.88$}
& \textcolor{red}{$2.54\pm0.36$}
& \textcolor{red}{$1.25\pm0.19$}
& \textcolor{red}{$1.67\pm0.48$}
& \textcolor{red}{$1.56\pm0.48$}
& \textcolor{red}{$0.854\pm0.181$} \\[0.15cm]

\multirow{2}{*}{17.2}

& $7.27\pm0.74$
& $3.08\pm0.28$
& $1.70\pm0.30$
& $1.62\pm0.38$
& $2.46\pm0.40$
& $1.08\pm0.19$ \\

& \textcolor{red}{$6.48\pm0.46$}
& \textcolor{red}{$2.92\pm0.40$}
& \textcolor{red}{$1.47\pm0.23$}
& \textcolor{red}{$1.52\pm0.13$}
& \textcolor{red}{$1.82\pm0.73$}
& \textcolor{red}{$0.974\pm0.218$} \\[0.15cm]

\bottomrule
\end{tabular}
\label{tab:b2bProx}
\end{table*}

\bigskip 
\noindent\textbf{Arterial Point Cloud Persistence.}
The alpha-complex analysis of the arterial point cloud revealed pressure-, \mbox{dimension-,} and lobe-dependent differences between CTRL and HPH mice.

In dimension 0, the strongest group separation occurred at 6.3 mmHg. Across pressures, HPH mice generally showed lower Betti-curve maxima in the left and inferior lobes. Except for the middle lobe, for the HPH mice the degree-zero lifespan curves (before normalization) had higher maxima compared to the controls. Exploratory comparisons also identified differences between the normalized lifespan curves at 6.3 mmHg. At 6.3 mmHg, permutation tests distinguished CTRL and HPH mice in all five lobes, with the $L^2$ distances being significant for the lifespan and normalized lifespan curves.

\begin{figure}[h!tbp]
    \centering
    \includegraphics[width=0.85\linewidth]{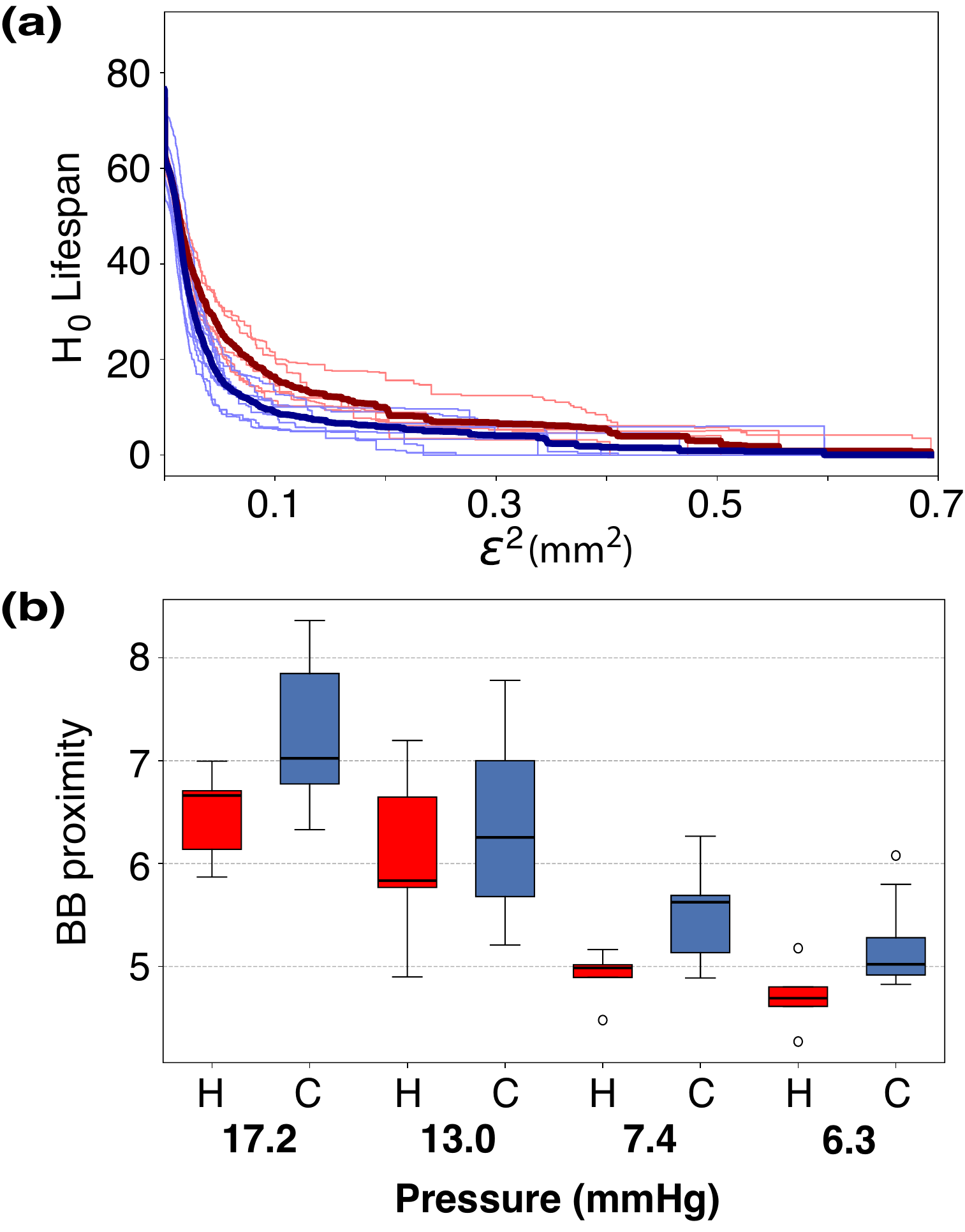}    \caption{Summary of TDA Results. (a) $H_0$ lifespan curves of the full network complement for CTRL (blue) and HPH (red) in dimension 0 at 7.4 mmHg. Dark lines are average curves, while the individual lines are plotted lighter. (b) Box-and-whisker plot showing the branch-to-branch (BB) proximity (the total finite $H_1$ persistence of the arterial point cloud) across all four pressures.}
    \label{fig:tdaResults}
\end{figure}

A particularly consistent pattern was observed in dimension 1. Across the full arterial network, total $H_1$ persistence, summarized by branch-to-branch proximity, was lower in HPH mice than in CTRL mice at all four pressures; see Figure~\ref{fig:tdaResults}(b).
The same pattern occurred in 19 of the 20 lobe--pressure combinations (Table~\ref{tab:b2bProx}). The reduction in total $H_1$ persistence was accompanied by lobe-dependent differences in the persistence curves. Total persistence depends on both the number of finite persistence intervals and their mean length. Distinguishing their contributions requires separate comparisons of interval counts and mean lifetimes calculated for each mouse.

The strongest $H_1$ curve separation occurred in the middle lobe at 7.4 mmHg, where the Betti, Lifespan, and Norm Lifespan curves differed between groups (\(p=0.001\)--\(0.018\)). At this pressure, the mean lifetime of $H_1$ classes was approximately $20\%$ lower in HPH mice. Additional significant differences occurred in the inferior lobe at 17.2 mmHg, where CTRL mice had higher $H_1$ peaks than the HPH mice ($p=0.005$ for the $L^2$ distance).

Dimension 2 contained substantially fewer and less persistent features. Qualitatively, Betti-curve maxima tended to be lower in HPH mice in the left and inferior lobes, whereas lifespan and normalized lifespan maxima were more often higher in HPH mice. The corresponding Betti values and persistence-landscape amplitudes were small relative to those in dimensions 0 and 1. However, localized group differences were detected, most notably in the inferior lobe at 17.2 mmHg, where $\lambda_2$ and $\lambda_3$ differed between groups ($p=0.004$ and $p=0.001$, respectively), with HPH mice exhibiting higher landscape amplitudes.

Overall, the arterial point-cloud analysis revealed the biggest $H_0$ differences at low pressure and a consistent reduction in total $H_1$ persistence in HPH mice. The $H_1$ persistence curves showed lobe-specific differences in the distribution of persistence across filtration scales, while the $H_2$ differences were more localized. Full permutation-test results are provided in Supplementary Materials Tables~S11 and S12.

\medskip
\noindent\textbf{Complement Persistence.} The persistence of the complement point cloud revealed pressure-, dimension-, and lobe-dependent differences between CTRL and HPH mice.

In dimension 0, the weighted HPH lifespan curves retain larger values over the later filtration range, see Figure \ref{fig:tdaResults}(a). The persistence landscapes showed a similar qualitative pattern, with higher HPH maxima in the full network and in several individual lobes. However, the full-network landscape differences were not statistically significant. At 7.4 mmHg, exploratory comparisons identified group differences in both lifespan and normalized lifespan curves. Visual inspection suggests that the unnormalized lifespan curves decayed slower in HPH mice. Thus, the $H_0$ differences were most apparent at the lower pressures, see Supplementary Materials Table S13.

The full-network complement analysis revealed differences between CTRL and HPH mice in dimension 1 at all four pressures. Betti and lifespan-type curves generally attained lower maxima in HPH mice, whereas the leading persistence landscapes tended to attain higher maxima and have broader support. At $7.4$ mmHg, the CTRL Betti, lifespan, and normalized lifespan curves appeared to decline more gradually after their maxima. Global permutation tests based on the $L^2$ distance between group mean curves identified differences between groups; each of the first three persistence landscapes also differed ($p=0.001$, $0.002$, and $0.015$ for $\lambda_1$, $\lambda_2$, and $\lambda_3$, respectively). At $13.0$mmHg, all three landscapes also differed ($p=0.012$, $0.019$, and $0.049$, respectively), with additional significant $H_1$ differences detected at $6.3$ and $17.2$ mmHg. 

Dimension 2 showed a pronounced lobe- and pressure-dependent pattern. Betti-curve maxima were generally lower in HPH mice, whereas the direction of the lifespan and persistence-landscape differences varied among lobes. At 6.3 and 7.4 mmHg, significant Betti-curve differences were found in the left, middle, inferior, and post-caval lobes. The inferior lobe showed particularly consistent landscape differences (HPH had higher peaks), with a significant dimension-2 persistence-landscape comparison at each pressure and the strongest separation at 6.3 mmHg ($\lambda_1$, $p=0.001$). 

Overall, the complement analysis showed the biggest full-network differences in $H_1$, while $H_2$ differences were more localized and particularly prominent at lower pressures. The inferior and left lobes exhibited the most extensive lobe-specific differences. Full permutation-test results are provided in Supplementary Materials Tables~S13 and S14.

\subsection{Statistical Shape Analysis}
\smallskip

\noindent The top 20 principal component scores from shape PCA account for $94-97\%$ of the variation in the landmark data, depending on which lobe the algorithm was applied to. Therefore, we restrict our use of the principal component scores to just the top 20. An ANCOVA test of these top 20 shape PCA scores shows that only the second principal component is significantly different between the hypertensive and control groups. Therefore, we fit an SVM model using only the first two principal components. The results of each lobe are given in the Supplementary Materials Table~S16. Despite the small sample size, we use these findings to explore the predictive potential of SSA features alone and to identify which pressures or lobes are best suited for using SSA features as a potential diagnostic for pulmonary hypertension. 

\begin{figure}
    \centering
    \includegraphics[width=0.85\linewidth]{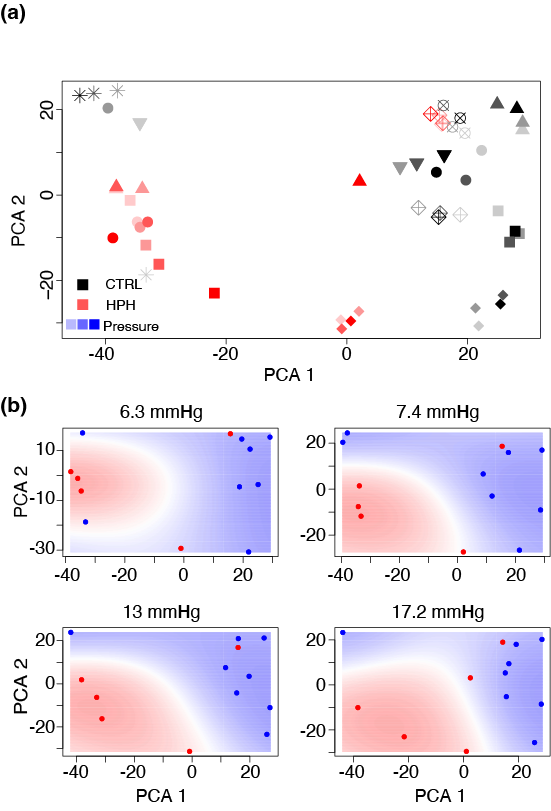}
    \caption{(a) Scatterplot of the first and second shape PCA scores. Each icon represents a different mouse, and different shades of the same icon represent different pressures. HPH mice generally have lower second principal component scores at 7.4, 13.0, and 17.2 mmHg. (b) Four SVM predictive models are fit on each pressure separately. The 7.4 mmHg and 13 mmHg pressures appear to separate control and hypertensive arteries best. }
    \label{zfig:SSAResults}
\end{figure}

The results for the full network are shown in Figure~\ref{zfig:SSAResults}. We use leave-one-out cross-validation to evaluate the SVM model's accuracy, which peaks at around 77\% at pressures of 7.4 mmHg and 13 mmHg. The accuracy for pressures 6.3 mmHg and 17.2 mmHg is 53.8\% and 69\%. While using the full network was generally the most accurate, the superior lobe's model performed similarly well, having higher accuracy than the other lobes when using the SSA features for prediction. Within the superior lobe, the lowest pressure (6.3 mmHg) had the highest accuracy (77\%), followed by 7.4 and 13.0 mmHg (69\%). The worst-performing pressure was 17.2 mmHg at 53.8\% accuracy. It intuitively makes sense that the pressure of the superior lobe is inversely related to the accuracy of the model, because the control and hypertensive pulmonary networks most resemble each other at the highest pressure. Across each of the pressures, we found that the HPH mice are characterized by lower scores of principal component 1. In the higher pressures of 7.4, 13.0, and 17.2 mmHg, the HPH mice tend to have lower second principal component scores as well. These  results suggest differences in shape between the hypertensive and control mice. 

\subsection{Classification of pulmonary hypertension using all Feature Groups}
\smallskip

\noindent The posterior inclusion probabilities of the full network's sparse linear regression model are provided in Table \ref{tab:importance_full}. For the full network, the most important feature was the standard deviation of $L_R$, followed by the maxima of the third landscape curve (dimension 2) and the Betti curve (dimension 2). After this, most other features were approximately the same posterior inclusion probabilities.  The results for the lobe-specific models are provided in the Supplementary Materials Section 2.4 Tables S17 and S18. 

The \texttt{gpboost} package reported the median of $\alpha$ as the most important feature for predicting pulmonary hypertension classification across all models at the full-network level and the two lobe-specific levels. At the full network level, these results are consistent with the sparse linear regression posterior inclusion probabilities. For the ``wide-lobe" level, which provides variables from each lobe independently, the median $\alpha$ from the middle lobe was the most important, followed by the standard deviation of $\alpha$ in the inferior lobe.

\begin{figure}
\centering
    \includegraphics[width=0.85\linewidth]{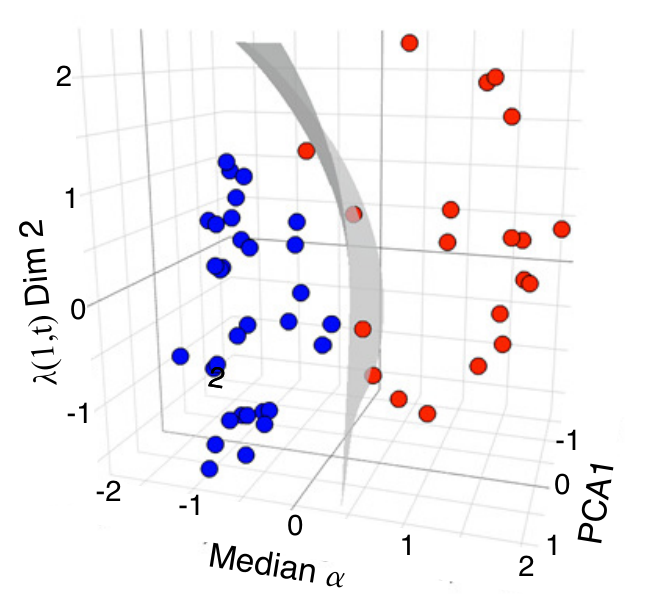}
    \caption{3D visualization using three predictors from each category (geometric, TDA, SSA) in an SVM model: the first shape principal component, the median of $\alpha$, and the active range of the 1st landscape in dimension 2 seem to be jointly informative of whether a mouse has pulmonary hypertension (red) or is a control (blue). Units for each of the features are standard deviations from their respective means. The LOOCV accuracy of this dataset is about 69.2\%, whereas using the entire dataset is around 71\% accurate. }
    \label{fig:hypertension_diagnosis}
\end{figure}

For the final SVM model fit and visualization, the full network level was chosen for its simplicity. Furthermore, we chose the median of $\alpha$, the first shape principal component, and the active range of the 1st landscape in dimension 2. A 3D visualization of the SVM model with its decision boundary is shown in Figure~\ref{fig:hypertension_diagnosis}. When using leave-one-out cross-validation, we achieve about 69\% accuracy using only these three variables. When using all the variables of the full network, we only marginally improve our LOOCV accuracy to about 71\%. 
 
\begin{table*}[ht]
\centering
\caption{Top 10 variable importance rankings for the sparse linear regression and GPBoost models for the full mouse network.}
\label{tab:importance_full}
\begin{tabular}{llll}
\toprule
\multicolumn{2}{c}{\textbf{Sparse Linear Regression Model}} &
\multicolumn{2}{c}{\textbf{GPBoost Model}} \\
\cmidrule(lr){1-2} \cmidrule(lr){3-4}
\textbf{Variable} & \textbf{Importance} &
\textbf{Variable} & \textbf{Importance} \\
\midrule
StDev $L_R$               & 0.1005 &
Median $\alpha$                      & 0.7385 \\
Max Landscape 3 Dim 2                & 0.0747 &
Median $\xi$                         & 0.1504 \\
Max Betti Curve Dim 2                & 0.0741 &
StDev $\beta$                        & 0.0259 \\
Max Landscape 1 Dim 1                & 0.0663 &
Median $\eta$                        & 0.0254 \\
Peak Lifespan Dim 1, C         & 0.0632 &
Max Landscape 5 Dim 1                & 0.0201 \\
Max Norm. Lifespan Dim 1, C     & 0.0618 &
$L_1$ Norm Landscape 1 Dim 1, C & 0.0186 \\
AUC Landscape 3 Dim 1                & 0.0609 &
20th Shape PCA Score                    & 0.0065 \\
StDev $\psi_1$                       & 0.0594 &
Max Landscape 6 Dim 1                & 0.0056 \\
AUC Landscape 3 Dim 2                & 0.0569 &
Range Landscape 1 Dim 1, C      & 0.0033 \\
11th Shape PCA Score                    & 0.0547 &
Standard Deviation $\alpha$             & 0.0030 \\
\bottomrule
\end{tabular}
\smallskip
\\
\textit{Note:} C = complement.
\end{table*}

\section{Discussion}
\noindent The objective of this study was to describe remodeling associated with group 3 pulmonary hypertension via analysis of micro-CT images of arterial networks. The study included analysis of pulmonary arterial networks extracted from control and hypoxia-induced pulmonary hypertension animals. Overall, the large arteries in the hypertensive animals have larger radii than those in the control animals (i.e., here, large arteries refer to arteries resolved in the micro-CT images). This finding agrees with our previous study \cite{Miller23} and several physiological studies, e.g.,~\cite{Chen2015, Hoesein2016,Chin2018}. 

While this finding is essential, our study adds additional metrics that differentiate remodeling across different parts of the lung. Discussion of each metric derived using geometric analysis, persistent homology, and statistical shape analysis is provided below.

\subsection{Geometric Features}
\smallskip

\noindent Most studies, including work by Chen et al. \cite{Chen2015}, Hoesein et al. \cite{Hoesein2016}, and Chin et al. \cite{Chin2018}, found that vessel radii in the main, left, and right pulmonary arteries were larger in subjects with pulmonary hypertension than in controls. Our study agrees with this finding but goes a step further by analyzing how network properties change as vessels get smaller. In this section, we compare features extracted from large pulmonary arterial networks in control and hypertensive animals. Our results showed that within the imaged region, the difference in length-to-radius ratio between the control and hypertensive networks remains constant. The increased vessel radius may be associated with arterial wall thickening, consistent with Chazova et al. \cite{Chazova95}, who reported that the pulmonary arterial wall is thickened and that thickening is greater in smaller vessels. This may be due to calcification and increased inflammation of the arterial wall, as suggested by Liu et al. \cite{Liu22}.

Several studies (e.g., \cite{Singh2022}) have found that the smallest vessels in patients with group 3 pulmonary hypertension have smaller radii and fewer vessels at the microvascular level compared to controls. Histopathological evidence explains this finding, as distal pulmonary vascular remodeling is associated with medial and intimal thickening, arteriolar muscularization, and increased luminal stenosis \cite{Igarashi2025}. These changes reduce the effective cross-sectional area of the distal pulmonary vascular bed and contribute to increased pulmonary vascular resistance. In our studies, we found that the radius ratio, relating the size of the daughter vessels to the parent vessel ($\alpha$), is greater in hypertensive animals. Meanwhile, the branching factor ($\beta$) remains the same.  Assuming the capillaries have approximately the same radii in control and hypertensive animals, a higher value of $\alpha$ correlates with a tree that bifurcates more times before reaching the capillary level. A higher value of $\alpha$, which, with an unchanged value of beta, produces tress that are more asymmetric with longer pathways. Flow along such trees will generate higher peripheral resistance. This finding also agrees with results from persistent homology measures discussed below.  

While the average value of $\alpha$ is higher in the hypertensive animals, the same conclusion cannot be made when comparing $\alpha$ and $\beta$ as a function of radius (shown in Figure~\ref{fig:geomFigures}); the results here show a clear variation with radius: the vessels get more symmetric as they get smaller, but there is no difference between the control and hypertensive animals. This finding may be explained by the small number of samples or the large variance in the data. One reason we did not find differences could be that the data were obtained from adult mice placed in hypoxic chambers, so the analysis was done before extensive remodeling developed. The animals examined in this study were placed in a hypoxic environment for 10 days \cite{Vanderpool11}, which, as shown in this and previous studies \cite{Cahill2012}, is long enough to generate pulmonary hypertension but may not be enough time to develop significant vascular remodeling. Additionally, we saw in Figure \ref{fig:geomFigures}(b) that at the smallest radii, $\alpha$ and $\beta$ tend toward 1, indicating that these bifurcations see barely any change in radius between parent and daughter vessels. This suggests that these smallest vessels exhibit different geometric properties than vessels with radii between 100-150 $\mu$m. Another possibility is that these vessels' radii are near to the image resolution, which can cause errors in extracting geometry.

Another way to examine the tree is to compare the asymmetry ratio of the smaller and larger daughter vessels, as defined by Zamir \cite{Zamir78}. In the hypertensive mice, we see significant decreases in the mean value for $\gamma$, aligning with the increases in $\alpha$ and relatively constant $\beta$ across groups. This finding aligns with the study by Olufsen et al. \cite{Olufsen2012}, which found that decreases in $\gamma$ lead to the rarefaction of distal vessels in the network. Antonios et al. \cite{Antonios03} suggest that rarefaction may cause pulmonary hypertension in the larger vessels. $\eta$ measures the ratio of cross-sectional areas between daughter and parent vessels. We find significant increases in this ratio, which also points towards rarefaction in the distal vessels \cite{Olufsen2012}. The last radius relation is known as Murray's exponent \cite{Murray26,Murray26a, Uylings77} ($\xi$) which depends on $\alpha$ and $\gamma$. We directly calculate $\xi$ via Newton's method to limit the variability introduced by the variance of other parameters. Altieri Correa et al. \cite{Correa25} performed a study on $\xi$ using three healthy mice and found the mean value to be approximately 2.6, but ranged from roughly $1.5$ to $3.4$. Our findings showed that the mean $\xi$ increased from $2.4$ to $2.9$ under pulmonary hypertension, which aligns with previous findings \cite{Correa25}. 

Each bifurcation has four angles able to be extracted. Though these angles do not differ in their distribution, they provide a new avenue for studying how disease affects vessel networks in any organ. Other features intrinsic to the arterial network are extracted, while not finding differences in this organ-disease system, may be insightful in other systems. Further, the extraction methodology from the segmentation of micro-CT images offers a noninvasive way to identify signs of pulmonary hypertension.

\subsection{Persistent Homology}
\smallskip
\noindent Persistent homology extends the geometric analysis by describing how HPH alters the organization of the pulmonary arterial network in three-dimensional space. Rather than indicating a uniform loss of topological complexity, the results suggest pressure-, direction-, and lobe-dependent remodeling. For the arterial centerline point cloud, the most significant $H_0$ differences occur at low pressure, particularly at 6.3 mmHg, where CTRL and HPH mice differ across all five lobes. Since $H_0$ persistence reflects the scales at which spatially separated portions of the arterial tree merge, these findings are consistent with altered spacing and packing of arterial branches. The prominence of this effect at low pressure may reflect differences in vascular mechanics that become less apparent as the vessels distend, consistent with the reduced compliance and structural remodeling associated with hypoxic pulmonary hypertension.

A particularly consistent finding is the reduction in total $H_1$ persistence, or branch-to-branch proximity, in HPH mice. This occurs across the full network at every pressure and in 19 of the 20 lobe--pressure combinations. A consistent descriptive finding is the lower mean total $H_1$ persistence, referred to as branch-to-branch proximity, in HPH mice across the full network at every pressure and in 19 of the 20 lobe–pressure combinations. Together with the lobe-specific persistence-curve differences, this finding is consistent with altered spatial relationships among branches. However, total persistence combines interval number and mean lifetime, and the present summaries do not establish their separate contributions to the reduction. In particular, a lower Betti-curve maximum indicates fewer simultaneously active classes, rather than necessarily fewer classes over the entire filtration. Comparisons of interval counts and mean lifetimes calculated for each mouse are needed to determine how these contributions vary among lobes. This lobe dependence is consistent with the heterogeneous structural remodeling observed in hypoxic pulmonary hypertension \cite{Chambers20}.

The complement provides an additional geometric representation of the  unresolved intervascular region through its boundary surface. Its most consistent whole-network differences occur in $H_1$, while $H_2$ differences are more localized and particularly prominent at lower pressures. These findings are biologically relevant because pulmonary hypertension prominently affects the distal pulmonary circulation, including vessels that fall below the resolution of the micro-CT images. Imaging and histological studies have similarly linked changes in the visible pulmonary vascular tree, including vascular pruning, with remodeling of small pulmonary arteries \cite{Synn21Pruning}. It should be noted, though, that the complement is only a geometric surrogate for the unresolved intervascular space; its topology may therefore contain an indirect signature of distal vascular remodeling. This interpretation is consistent with the complement-based topological framework of Belchi et al.~\cite{Brodzki18}.

Moreover, the lower Betti and lifespan maxima, together with higher leading landscape maxima, are consistent with changes in the multiscale organization of the sampled vessel surfaces and enclosing boundary. This combination can arise when fewer features are simultaneously active, while some features persist over a wider range of filtration scales. Biologically, such a pattern could reflect remodeling of vessel caliber and spatial arrangement, altering the geometry of the surrounding intervascular space. However, the persistence summaries alone do not establish distal vessel loss or identify a specific remodeling mechanism.

The height filtration further shows that HPH alters the directional organization of the arterial tree. Directional complexity is lower on average in HPH mice in both anterior and posterior directions, but the number of finite $H_0$ intervals does not always decrease correspondingly. Thus, the directional difference cannot be explained by feature number alone; the persistence of those features is also altered. The additional lobe-specific persistence-curve and landscape differences are consistent with the direction-dependent changes previously reported by Miller et al.~\cite{Miller23}. Taken together, the alpha-complex, complement, and height-filtration analyses indicate that HPH reorganizes arterial spacing, branch-to-branch relationships, and directional structure in ways that are anatomically heterogeneous and are not captured by conventional geometric measurements alone.

\subsection{Statistical Shape Analysis}
\smallskip 

\noindent The results of the statistical shape analysis agree with previous findings in the field. In an earlier study from 2015, \cite{chan_quantifying} reports that a study using PCA with statistical shape analysis identified that most of the variability in the shape difference in human pulmonary arteries is the ``angles of fissures" and the difference in relative lobe size, both of which would affect the shape of the entire network. In a more recent study \citep{Sabry23}, shape PCA results revealed significant shape changes in human principal pathways--specifically the angles between the right and left principal pathways, as well as the root's diameter and bulging. Each of these would affect the overall shape of the convex hull of the entire network, as we have analyzed in this study. Our results imply that the superior lobe shape could differ systematically at lower pressures between the HPH and control groups. There also seems to be a difference in shape between the full networks near the pressures $7.4$ and $13.0$ mmHg.

The classification results suggest that the effect of pressure on vascular shape is not uniform across lobes. In general, lower pressures may be expected to yield greater separation between control and hypertensive networks. Control animals have more compliant arteries; as pressure increases, their arteries dilate more than those from hypertensive animals. Since the deformation is nonlinear \cite{Lee2016}, at the highest pressure the arteries of the control animals become morphologically more similar to hypertensive vessels. The superior lobe is broadly consistent with this interpretation, with the highest classification accuracy occurring at $6.3$ mmHg and decreasing at higher pressures. However, this pattern is not universal. Most notably, the full-network analysis achieves its highest accuracies at $7.4$ and $13.0$ mmHg rather than at the lowest pressure, suggesting that the geometry of the entire arterial network may contain discriminative shape information that becomes more apparent at intermediate pressures. The remaining individual lobes show weaker classification performance across pressures, with accuracies generally near 50\%. This may indicate that, for these lobes, pulmonary hypertension does not produce a sufficiently consistent structural change in the convex hull geometry to distinguish the two groups using SSA alone. Again, this finding may be related to the experimental setting in which the data analyzed here were obtained.

The statistical shape and topological analyses can also be viewed as complementary descriptions of pulmonary arterial morphology. SSA characterizes the shape of the network’s convex hull, while the boundary-point-cloud persistence analysis characterizes multiscale features of the sampled hull and modeled vessel surfaces. This distinction may be particularly relevant to pulmonary vascular remodeling. For example, \citep{Synn21Pruning} found that increased remodeling of the small pulmonary arteries was associated with greater vascular pruning, reflecting a loss of small-vessel volume that may alter the distribution of empty space within the overall extent of the arterial network. Similarly, \citep{Sun2011Fractal} demonstrated that pulmonary hypertension was associated with differences in both the fractal dimension and projected area of the pulmonary arterial tree, suggesting that disease-related changes occur not only in the overall shape of the vasculature, but also in how the vessels fill the surrounding space. The convex hull captures changes in the overall geometry of the network, while the complementary TDA curves capture changes in the organization and occupancy of the vasculature within that geometry.

\subsection{Classification of pulmonary hypertension}
\smallskip
\noindent The multi-feature classification analysis demonstrates that integrating distinct feature groups--specifically geometric features, topological data analysis (TDA), and statistical shape analysis (SSA)--yields a robust framework for detecting pulmonary hypertension. By first identifying active effects using Bayesian sparse linear regression and important features from boosted trees, we leveraged a combined feature set to achieve reasonable cross-validation accuracy while drastically reducing dimensionality. Combining these feature spaces ensures that organ deformation, local vessel geometry, and topological network remodeling are jointly accounted for, providing a comprehensive set of features for identifying hypertension using features derived solely from network morphology.

Previous studies have also demonstrated that quantitative pulmonary vascular morphology contains relevant information for diagnosing pulmonary hypertension. Sun et al.~\cite{Sun2011Fractal} used fractal dimension analysis on human networks to diagnose pulmonary hypertension. Rahaghi et al.~\cite{Rahaghi16, RAHAGHI2021} found differences in vascular pruning, tortuosity, and vessel volume in patients with pulmonary hypertension. Most recently, \cite{Beste2026diagnostics16040619} found that patients with pulmonary hypertension had relatively more 6--10 mm arteries. Our analysis extends this morphology-based paradigm by incorporating features from three distinct categories: vessel geometry, shape, and topology, which could be used as a stepping stone for future non-invasive diagnostics for pulmonary hypertension in humans. In our findings, $\alpha$ was found to be the most important predictor according to the GP Boost package in R, which was consistently found to be larger in HPH mice than control mice in Section \ref{sec:geom}. Although $\alpha$ was identified as the most important predictor by GPBoost, this result should be interpreted in the context of the classification objective. It is possible that $\alpha$ captures information shared with other predictors. Consequently, the model may favor $\alpha$ without assigning substantial additional importance to these correlated features. Therefore, although $\alpha$ was the primary predictor used in the classification model, this does not necessarily contradict the results obtained from our other analyses.

\subsection{Limitations}
\noindent The primary limitation of this study is the small dataset; we analyzed networks from only eight control and five hypertensive mice. The small sample size makes the classification task more susceptible to random noise. It also interacts with our validation scheme: with leave-one-out cross-validation, the training set for these models is biased toward control mice; therefore, many classification errors were Type II, failing to identify hypertension.

A second limitation concerns the lobe boundary. Because we analyzed excised arterial trees, we could not directly access the lobe boundary, and for both the persistent homology and SSA analyses we used the convex hull as a surrogate. Because the landmarks that SSA extracts are defined along the convex hull rather than the centerlines, the resulting principal components primarily track volumetric expansion and changes in spatial reorientation between distinct lobes. Consequently, this approach cannot directly detect fine-scale remodeling of internal vascular branching patterns or vessel-level morphometric shifts within the pulmonary tree. Persistent homology provides additional descriptors of the sampled spatial geometry and directional organization of the arterial network. Dimension one and two features in the alpha filtrations describe tunnels and cavities of the filtered complexes, rather than anatomical loops or directly observed cavities in the arterial tree.

The TDA analyses are also limited by data preparation and by how we represented and compared topological information. At lower pressures, HPH mice have considerably more vessels on the micro-CT image. This study used pruned networks to control for this difference; the observed topology therefore reflects both vascular biology and the image-to-network extraction pipeline. In addition, the classification models relied on summary statistics, rather than the full persistence curves, potentially overlooking differences in shape and the distribution of topological features across filtration. Furthermore, the permutation tests were performed separately at each pressure, unlike the ANCOVA tests used in the geometric feature analysis. The complement point cloud is sensitive to sampling density, because we sample a fixed number of points from around each vessel, the spacing between points increases with vessel radius.

Finally, a large number of statistical comparisons were performed throughout the study. We did not adjust the p-values for multiple comparisons across pressures or lobes. Given the exploratory nature of these analyses, we aimed to identify potentially informative morphological differences rather than provide confirmatory evidence for individual effects. Consequently, the reported p-values should be interpreted with caution, as the lack of multiple-testing adjustments increases the likelihood of false-positive findings.

\subsection{Conclusion}
\noindent This study provides a framework that includes a range of distinct yet complementary tools to study diseases in arterial networks in any organ. Starting with micro-CT scan segmentation, we extract arterial trees and convex hulls. Geometric features, including a new algorithm for calculating angles, identify bifurcation-level differences resulting from hypertension. Persistent homology and the statistical shape analysis extend these findings to overall morphological differences in the networks and can identify specific lobes in which morphological changes are more pronounced.

This pipeline highlights differences in radius relations, including $\alpha$, $\eta$, $\gamma$, $\xi$ and $\ell_{rr}$. The increase in $\alpha$ and $\eta$ and decrease in $\gamma$, along with a decrease in $\ell_{rr}$ with hypertension, consistent with the radius of the larger daughter vessel increasing, suggests that the conduit vessels expand, e.g., to prevent excessive increase in peripheral vascular resistance. The persistent homology analysis shows that branch-to-branch proximity and directional complexity decrease with disease, possibly suggesting the trees "rarefy," a pattern consistent with hypertension. Finally, Betti and lifespan curves reveal lobe-specific changes in dimensions 0 and 2 in the left and inferior lobes; the small sample size did not allow us to differentiate how each lobe remodeled, but remodeling is likely heterogeneous, i.e., different animals may be affected in different orders. The shape statistics analysis also demonstrated that the control and hypertensive mice may differ in shape. Lastly, an SVM predictive model was constructed using the median $\alpha$, first principal component on the shape statistics, and the first landscape in dimension 2, to separate the groups with 71\% accuracy.

The analysis used here has the potential to help differentiate remodeling and be coupled with studies examining mechanisms of remodeling. Differences observed can also be incorporated into network-generation algorithms, such as constrained constructive optimization, to create disease-specific network models. Overall, this framework offers an adaptable pipeline that can be applied across organs to study vascular remodeling in disease. 

In summary, the three analysis methods all show that the arterial tree remodels with hypertension. Geometry detects altered vessel caliber and bifurcation scaling; TDA detects reduced branch-to-branch and directional persistence together with lobe- and pressure-specific changes; and SSA detects disease-associated changes in overall network and lobe shape. The classification analysis explores the joint predictive use of features from all three domains; whether each domain contributes additional predictive information remains to be established.

\subsection*{Supplementary information}
\medskip
\noindent Supplementary Tables and a detailed explanation of alpha complexes are available in the Supplement. Code, including arterial trees and convex hulls extracted from the CT images, tables of birth and death values extracted from the graphs, and code needed to regenerate the analysis in this study, can be found here: 10.5281/zenodo.22287168

\subsection*{Acknowledgements}
\smallskip 

\noindent This work was supported in part by the National Science Foundation, via the DRUMS REU award NSF-DMS \#2349611 and the UQ4Life RTG award NSF-DMS \#2342344. J Miller was supported by a NSF-GRFP award NSF-DMS \#1937954 and MS Olufsen was supported in part by the NIH-NHLBI \#R01HL147590.  We would like to thank Dr. Naomi Chesler and Sofia Altieri Correa, Department of Bioengineering, UC Irvine, for sharing CT images. We would also like to thank Drs. Jay Mackenzie and Namshad Thekkethil, School of Mathematics and Statistics, University of Glasgow, UK. Finally, we would like to thank Pablo Hernandez, Vasomaly, for help with installing and running SGEXT, used for skeletonization of the CT images. 

\subsection*{Declarations}
\smallskip

\noindent The authors declare no competing interests.  

\subsection*{Ethics}
\smallskip

\noindent This manuscript consists of prospective analysis of existing micro-CT images made available by Naomi Chesler, Department of Biomedical Engineering, University of California Irvine. Data were collected at University of Wisconsin, Madison, and as stated in the previously published study by Vanderpool et al. (2011) \cite{Vanderpool11}.  All protocols and procedures were approved by the University of Wisconsin and Zablocki VA Medical Center Institutional Animal Care and Use Committees. This study received de-identified images, which were segmented. Computer code as well as outputs, and arterial networks generated from the data are all made available on Zenodo link (10.5281/zenodo.22287168) referenced in our manuscript.

\printcredits

\bibliographystyle{elsarticle-num}

\bibliography{cas-refs}

\end{document}